\documentclass[twocolumn,english,amsmath,amssymb,superscriptaddress,nofootinbib]{revtex4-1}
\usepackage[T1]{fontenc}
\usepackage[latin9]{inputenc}
\usepackage{amsmath}
\usepackage{amssymb}
\usepackage{graphicx}

\usepackage{color}

\usepackage{ulem}



\usepackage{babel}
\begin{document}

\title{Superradiant Masing with Solid-state Spins at Room Temperature}
\author{Qilong Wu}
\address{Henan Key Laboratory of Diamond Optoelectronic Materials and Devices, Key Laboratory of Material Physics Ministry of Education, School of Physics and Microelectronics, Zhengzhou University, Daxue Road 75, Zhengzhou 450052 China }

\author{Yuan Zhang}
\email{yzhuaudipc@zzu.edu.cn}
\address{Henan Key Laboratory of Diamond Optoelectronic Materials and Devices, Key Laboratory of Material Physics Ministry of Education, School of Physics and Microelectronics, Zhengzhou University, Daxue Road 75, Zhengzhou 450052 China }

\author{Hao Wu}
\address{Center for Quantum Technology Research and Key Laboratory of Advanced Optoelectronic Quantum Architecture and Measurements, School of Physics, Beijing Institute of Technology, Beijing 100081, China}
\address{Beijing Academy of Quantum Information Sciences, Beijing 100193, China}

\author{Shi-Lei Su}
\address{Henan Key Laboratory of Diamond Optoelectronic Materials and Devices, Key Laboratory of Material Physics Ministry of Education, School of Physics and Microelectronics, Zhengzhou University, Daxue Road 75, Zhengzhou 450052 China }

\author{Kai-Kai Liu}
\address{Henan Key Laboratory of Diamond Optoelectronic Materials and Devices, Key Laboratory of Material Physics Ministry of Education, School of Physics and Microelectronics, Zhengzhou University, Daxue Road 75, Zhengzhou 450052 China }

\author{Mark Oxborrow}
\address{Department of Materials, Imperial College London, South Kensington SW7 2AZ, London, United Kingdom}

\author{Chongxin Shan}
\email{cxshan@zzu.edu.cn}
\address{Henan Key Laboratory of Diamond Optoelectronic Materials and Devices, Key Laboratory of Material Physics Ministry of Education, School of Physics and Microelectronics, Zhengzhou University, Daxue Road 75, Zhengzhou 450052 China }
\author{Klaus M{\o}lmer}
\email{moelmer@phys.au.dk}
\address{Niels Bohr Institute, University of Copenhagen, 2100 Copenhagen,Denmark}

\begin{abstract}
Steady-state superradiance and superradiant lasing attract significant attentions in the field of optical lattice clocks, but have not been achieved so far due to the technical challenges and atom loss problem. In this article, we propose that their counter-part may be observed in the microwave domain with solid-state spins-microwave resonator systems at room temperature with realistic technical restrictions. To validate our proposal, we investigate systematically the system dynamics and steady-state by solving  quantum  master equations for the multi-level and multi-process dynamic of trillions of spins. To this end, we employ a mean-field approach, and convert the mean-field dynamics of the spin ensemble into the one in a more intuitive Dicke state picture. Our calculations show that for systems with nitrogen vacancy center spins and pentacene molecular spins the superradiant Rabi oscillations occur firstly due to transitions among different Dicke states, and the subsequent continuous-wave superradiant masing can achieve a linewidth well below millihertz. Our work may guide further exploration of transient and steady-state superradiant masing with the mentioned and other solid-state spins systems, such as silicon vacancy centers in silicon carbide and boron vacancy centers in hexagonal boron nitride, where the coherent radiation with ultra-narrow linewidth may find applications in deep-space communications, radio astronomy and high-precision metrology.  
\end{abstract}
\maketitle

\section{Introduction}

Superradiance was introduced in 1954 by Robert. H. Dicke as collective spontaneous emission of an atomic ensemble \citep{RHDicke1954}, and was then investigated in 1980s and thereafter. Initially, the superradiance was considered as a transient phenomenon caused by collective decay of excited atoms or molecules \citep{AVAndreev1980}, while in 2009, D. Meiser et al. proposed that steady-state superradiance \citep{DMeiser2009} can be achieved by compensating the collective decay with incoherent atomic pumping, and predicted a coherent radiation with millihertz linewidth for optical lattice clock systems. Since such ultra-narrow radiation may have applications in quantum metrology~\citep{ADLudlow}, there were rapid developments later both in theories, e.g. coexistence of superradiance and stimulated emission (termed as superradiant lasing) \citep{DATieri2017,KDebnath2018,YZhang2021}, and in experiments, e.g. proof-of-concept of the steady-state superradiance with rubidium Raman transitions \citep{JDBohnet2012}, superradiant pulses and Rabi oscillations with strontium clock transitions \citep{MANorcia2016,MANorcia2016-1}, as well as the superradiance pulses-based frequency measurement \citep{MANorcia2018,YZhang3}. However, because the optical lattice clock systems require ultra-high vacuum, ultra-low temperature and also suffer from atom loss, the true steady-state superradiance and superradiant lasing have not been demonstrated yet. 

\begin{figure}[!htp]
\begin{centering}
\includegraphics[scale=0.32]{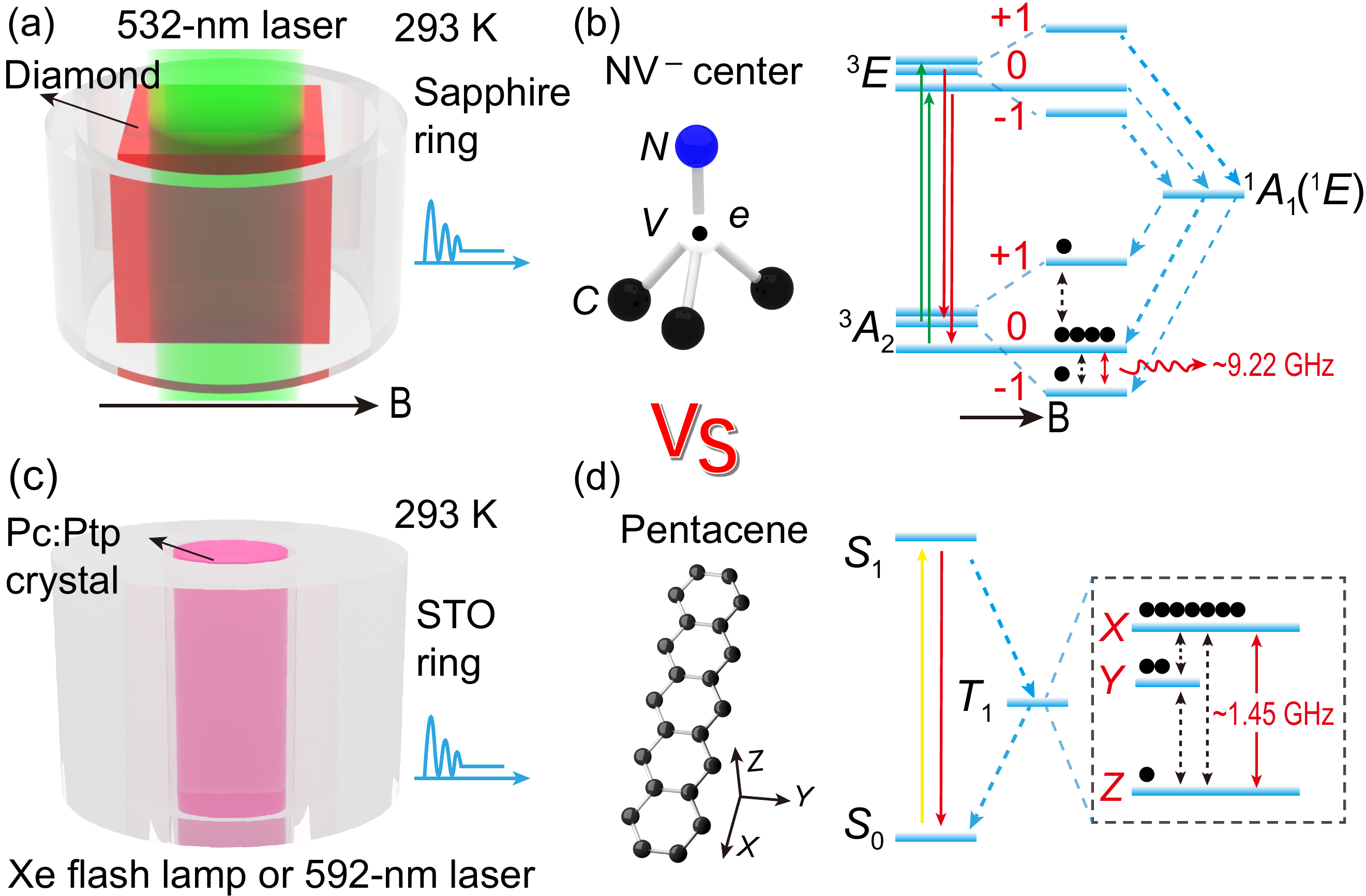}
\par\end{centering}
\caption{\label{fig:sys} Solid-state spins-microwave resonator systems at room temperature. Panel (a) shows a diamond excited by $532$ nm laser inside a sapphire ring in the presence of a static magnetic field $B$, as explored in the experiment \citep{Breeze}, while panel (b) shows the atomistic structure of the nitrogen-vacancy (NV) center, and the multiple electronic-spin levels and optical pumping processes among them (right). Panel (c) shows a pentacene-doped p-terphenyl crystal excited by a laser pulse or a Xenon flash lamp in a strontium titanate (STO) hollow cylinder, as explored in the experiments~\citep{JDBreeze,HWu1}, while panel (d) shows the atomistic structure of the pentacene molecules (left), and the essential electronic-spin levels and the optical pumping processes among these levels (right). Optical pumping in the NV centers and pentacene molecules introduce population inversion between the spin levels (black circles), which couple resonantly to the microwave resonators (red arrows). The blue curves on the right of the panel (a) and (c) show the initial superradiant Rabi oscillations and the subsequent continuous-wave superradiant masing. }
\end{figure}

In parallel with explorations in the optical domain, there was also intensive research of superradiance in the microwave domain with solid-state spins. The pentacene molecules \citep{TSLin} and nitrogen-vacancy (NV) centers in diamond \citep{BarryJF} received considerable attentions, because their spin levels have long coherence time at room temperature, and can be conveniently initialized and readout by optical means. They both were initially explored in optical detection magnetic resonance (ODMR) experiments \citep{Kohler,JWrachtrup1993,AGruber}, and were coupled with microwave resonators in recent years to realize pulsed/continuous-wave masing \citep{Oxborrow,HWu1,Breeze}, superradiant pulses \citep{ESalvadori,AAngerer2018}, Rabi oscillations \citep{JDBreeze,SPutz} and splittings \citep{RAmsuss,AAngerer}. Most of these experiments were achieved at room and cryogenic temperature for systems with pentacene molecules and NV centers, respectively, while recent theoretical \citep{YZhang2022PRL} and experimental \citep{DPFahey} work indicate the possibility of realizing them with NV centers also at room temperature. In view of these progresses, we propose that the solid-state spins-microwave resonator systems can be explored for the counterparts of steady-state superradiance and superradiant lasing in the microwave domain i.e. superradiant masing, with realistic technical constraints, see Fig. \ref{fig:sys}. The studies of these systems can deepen our understanding on the aforementioned phenomena, and help potentially the study of similar mechanics in optical domain.  

Despite the experimental progress, the dynamics and the steady-state of solid-state spins-microwave resonator systems are still not fully explored because of underlying multi-level, multi process dynamic of trillions of spins. More precisely, the optical spin polarization involves various processes among multiple electronic-spin levels (e.g. optical pumping, radiative decay, inter-system crossing and so on), and more than trillions of pentacene molecules/NV centers have to couple collectively and coherently with the microwave mode to achieve the desired phenomena. Although a quantum master equation can be easily established to describe the dynamics of such multi-level multi-particle multi-process systems, it can not be solved with standard density matrix technique. 

In this article, we leverage the mean-field methodology \citep{DPl}, initially developed to investigate the steady-state superradiance in optical domain \citep{DMeiser2009,CHotter}, to reduce the computation complexity significantly. More precisely, we consider the equations for mean-field quantities of direct interest instead of the full density matrix, and we apply the second-order cumulant expansion approximation, and assume all the pentacene molecules/NV centers identical to further simplify the equations to hundreds of coupled nonlinear differential equations. Note in such a treatment, we still preserve the collective nature of the spins-microwave mode coupling, and the core physics in an average sense. Furthermore, we apply the quantum regression theorem \citep{PMeystre} to calculate the radiation spectrum, and  provide also the formulas~\citep{KDebnath2018} to convert the mean-field quantities into the one within a more intuitive Dicke state picture~\citep{RHDicke1954}.

With these numerical tools, we are able to simulate more than trillions of optically pumped NV centers or pentacene molecules coupled with microwave resonators at room temperature, and analyze the system dynamics and steady-state properties. Our simulations show that superradiant Rabi oscillations occur firstly due to collective transition dynamics between different Dicke states, and the oscillation frequency is in the kHz and MHz range for the systems with NV centers and pentacene molecules, respectively. The subsequent continuous-wave superradiant masing has an extremely narrowing linewidth well below millihertz because of the exploration of sub-radiative Dicke states by the spin ensemble. The calculated cavity pulling factor, characterizing the variation of the superradiant masing frequency due to the frequency detuning between the relevant spin transition and the microwave resonator, is one order of magnitude smaller than unity, which signals the involvement of the spin superradiance but also the significant contribution from the stimulated emission. This analysis shows that the magnetic field fluctuations and the mechanical stability of the resonator must be well controlled to resolve and exploit the ultra-narrow radiation, otherwise the Zeeman-shift of fluctuated magnetic field and the frequency shift of the resonator will have an effect on it. All in all, our work provides physical insights into the transient and steady-state superradiant masing, and can guide the exploration of these phenomena with the mentioned and other solid-state spin systems, such as silicon vacancy centers in silicon carbide~\citep{HKraus,MFischer} and boron vacancy centers in hexagonal boron nitride~\citep{AGottscholl2020,AGottscholl2021}, where the promising ultra-narrow linewidth feature might find applications in deep-space communications, radio astronomy and high-precision metrology. 

The article is organized as follows. We explain the details of the considered solid-states spins-microwave resonators systems in the following section, and present the established quantum master equations to describe the dynamics of these systems and their solutions in the mean-field approach in Sec. ~\ref{sec:qme}. In Sec.~\ref{sec:rabi} and Sec.~\ref{sec:masing}, we discuss the results on the superradiant Rabi oscillations and the continuous-wave superradiant masing, followed by the analysis of the cavity pulling effect in Sec.~\ref{sec:pulling}. In the end, we summarize our work and comment on the possible studies in future.


\section{Solid-state Spins-microwave Resonators Systems}

In this section, we describe the solid-state spins-microwave resonator systems with the NV centers and the pentacene molecules, see Fig. \ref{fig:sys}. As shown in Fig. \ref{fig:sys} (a), the diamond with high NV concentration is illuminated by $532$ nm laser light inside a sapphire ring (high-Q microwave resonator) in the presence of a strong magnetic field. Fig. \ref{fig:sys} (b) shows the structure of the NV center (left part), i.e. a negatively charged point defect in the diamond lattice with a nitrogen atom replacing a carbon atom and a vacancy in the adjacent position, and the energy diagram of the NV center (right part), which contains a triplet ground state $^{3}A_{2}$, excited state $^{3}E$, and two singlet excited states $^{1}A_{1}, ^{1}E$ (represented by single meta-stable state). In addition, the triplet states contain three spin states with projection numbers $m_s = 0,\pm 1$ along the nitrogen-vacancy quantization axis, where the latter $m_s = \pm 1$ spin states are higher than the $m_s = 0$ state in energy due to the electronic spin-spin interaction. 

The NV centers can be optically pumped to the $^{3}E$ state, and then decay radiatively back to the $^{3}A_{2}$ state, or non-radiatively and spin-sensitively through the meta-stable excited states, leading to the higher population on the $m_s = 0$ spin level. This process was conventionally explored in ODMR experiments ~\citep{AGruber}, and was recently explored to cool the NV spin ensemble and to further cool the coupled microwave mode~\citep{WNg,DPFahey}, or to realize cavity-quantum electrodynamics effects at room temperature~\citep{YZhang2022PRL,YZhang1}. Furthermore, the Zeeman effect introduced by the applied magnetic field shifts the $m_s = -1,+1$ spin levels downwards and upwards, respectively, which can lower the $m_s = -1 $ spin level even below the $ m_s = 0 $ spin level, leading to a population inversion between these levels. In the previous experiment ~\citep{Breeze}, the population-inverted spin ensemble was thus coupled to a high-Q microwave mode of the sapphire ring to realize continuous-wave masing. In this article, we reveal the unexplored transient dynamics and steady-state properties, such as the superradiant Rabi oscillations and ultra-narrow linewidth, their dependence on the applied laser power, as well as the influence of the spin-resonator frequency detuning. 


Fig. \ref{fig:sys} (c) shows the system with pentacene molecules, as investigated in the experiment \citep{JDBreeze,HWu1}. This system consists of the p-terphenyl crystal with highly-doped pentacene molecules, illuminated by a laser pulse~\citep{JDBreeze} or a Xenon flash lamp~\citep{HWu1}, placed inside a STO ring (as a microwave resonator). Fig. \ref{fig:sys} (d) shows the structure of the pentacene molecule with five benzene rings (left part), and the simplified energy diagram (right part), which includes a singlet ground state $S_0$, an excited state $S_1$ and a lowest triplet state $T_1$. The $T_1$ state has three spin states with quantization axes along the long $X$, short in-plane $Y$ and out-of-plane $Z$ axis of pentacene molecule. The pentacene molecules can be optically excited from the $S_0$ state to the $S_1$ state, and then decay radiatively back to the $S_0$ state, or non-radiatively to the $T_1$ state via the inter-system crossing and internal conversion (through several higher triplet excited states~\citep{HWu2019}, not shown). Because the inter-system crossing rate of populating the $X$ spin state is much faster than that of the $Z$ spin state, we can create strong population inversion between these states, and then couple them to a high-Q microwave resonator to realize various effects. 

Since the metastable triplet states decay also rapidly to the $S_0$ state, it was initially believed that the pentacene molecules system can support only pulsed phenomena, such as the pulsed masing~\citep{Oxborrow} and Rabi oscillations~\citep{JDBreeze}. However, according to the more precise estimation of the spin level decay and dephasing rates ~\citep{HWu2019}, it was demonstrated in recent experiments~\citep{HWu1} that this system can actually also support the continuous-wave (CW) masing, for which, however, the underlying mechanism is still not fully characterized. In comparison to the earlier studies, in this article, we reveal that the superradiant Rabi oscillations originate from the collective dynamics in the Dicke state space, and predict the laser power dependence of the continuous-wave masing intensity and its linewidth.

\section{Quantum Master Equations and Mean-field Solutions \label{sec:qme}}

In this section, we describe the quantum master equations and the mean-field solutions, to address the dynamics and the steady-state of the solid-state spins-microwave resonator systems. To model theoretically the processes involved in the system with NV centers, we modify the quantum master equation for density operator as developed in our previous work for other effects~\citep{YZhang1}, and treat the NV centers as seven-levels systems to account for the electronic-spin levels and the microwave resonator as a quantized harmonic oscillator, and we account for various processes, i.e. the optical pumping, the radiative decays, the inter-system crossings, the spin-lattice relaxation, the spin dephasing, and the intra-resonator photon loss, as well as the collective and coherent NV spins-microwave photon energy exchange, see also Appendix \ref{sec:Master-NV}.

To simulate the system with trillions of NV centers, we can not rely on the standard density matrix technique due to the large number of matrix elements involved. Instead, we utilize the cumulant mean-field approach in second-order and also assume identical conditions for the NV centers \citep{QWu2021,YZhang1}, which together reduce dramatically the number of coupled equations, see Appendix \ref{sec:Master-NV}. As a result, we obtain tens of coupled differential equations for first-order mean-field quantities, such as the spin levels populations of the NV centers, and the second-order mean-field quantities, such as the mean photon number and the quantum correlations between the representative NV center pairs, where the latter are essential to capture the collective effects.

To analyze the mechanism leading to the superradiant Rabi oscillations and the continuous-wave superradiant masing, we employ the Dicke state picture to illustrate the quantum states of the NV spin ensemble. The Dicke states $\left | J, M \right \rangle$ with integer or half-integers $J\le N/2$ and $-J\le M\le J$ for two-level emitters ensemble are convenient to illustrate the collective coupling with the radiation field. Here, the number $M$ indicates the excitation degree of the ensemble, and the number $J$ indicates the symmetry of states or the strength of coupling (for more detailed information see \citep{YZhang4,QWu2021}). The Dicke states for given $M$ are usually visualized as ladders with equal spacing, and the states for different $J$ are shown as shifted ladders, forming a triangle space [inset of Fig. \ref{fig:rabi}(b)]. In  Appendix \ref{sec:Master-NV}, we provide the expressions to convert the mean-field quantities for multi-level systems to the averages of the Dicke state quantum numbers, which are associated with the two spin levels coupled resonantly to the microwave resonator.  

To simulate the processes in the system with pentacene molecules, we follow the same procedures by treating the pentacene molecules as five-levels systems, accounting for various processes among these levels, and solving the quantum master equation with the mean-field approach, see  Appendix \ref{sec:Master-NV}.

\begin{figure*}[!htp]
\begin{centering}
\includegraphics[scale=0.37]{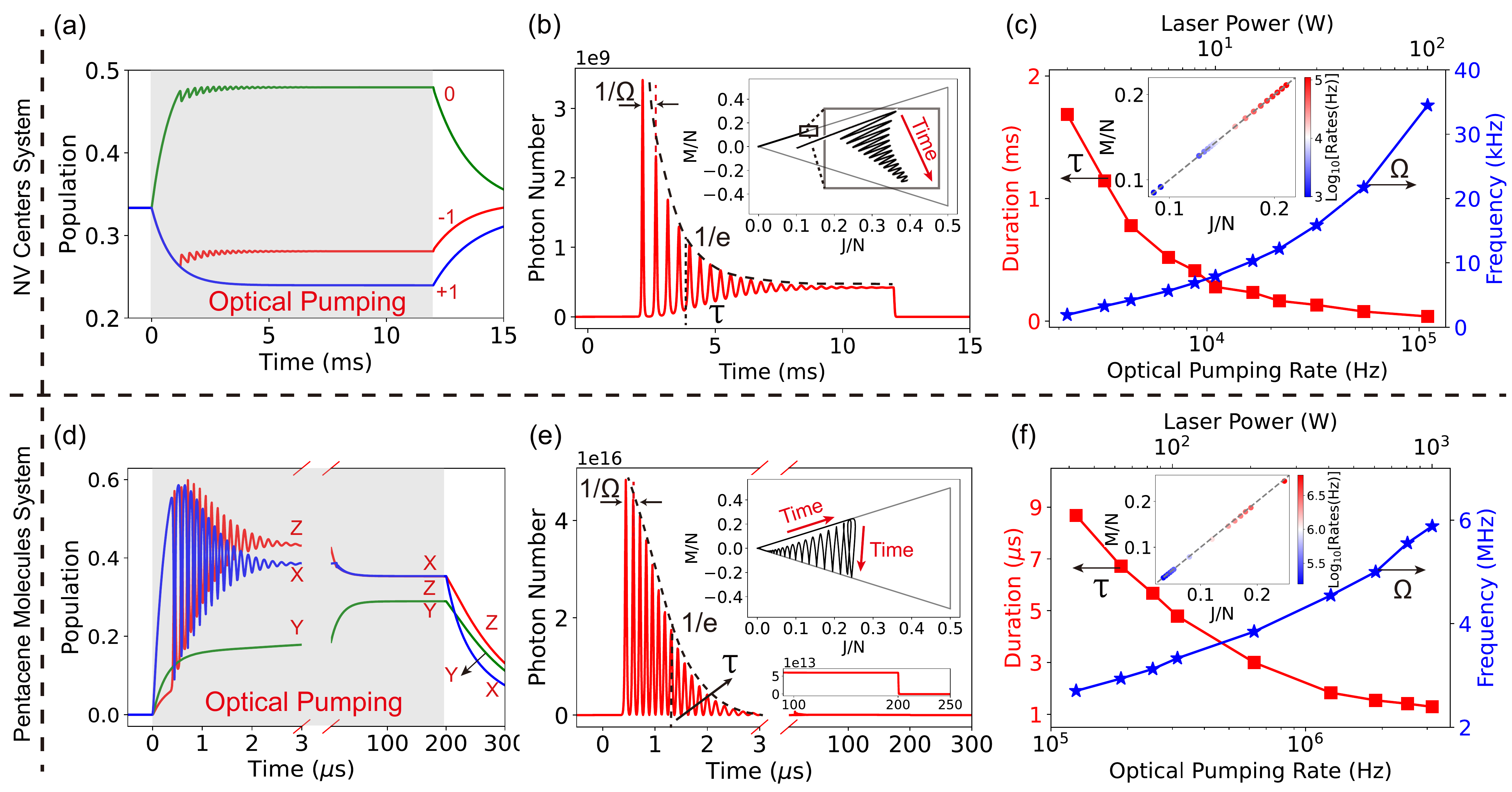}
\par\end{centering}
\caption{\label{fig:rabi} Superradiant Rabi oscillations and continuous-wave superradiant masing in systems with NV centers (a-c) and pentacene molecules (d-f) at room temperature. Panel (a,b) show the dynamics of the population of the electronic ground state spin levels (a), the NV spin ensemble in the Dicke state space [inset of the panel (b)], and the intra-resonator photon number (b), for the system excited by a $532$ nm laser pulse~\citep{WNg} with $2$ W power and $12$ ms duration. Panel (c) shows the oscillation duration $\tau$ (left axis) and oscillation frequency $\Omega$ (right axis) as a function of the optical pumping rate (lower axis) or the laser power (upper axis), where the value up to $20$ W might be achieved in the experiments ~\citep{ASarkar}, and the Dicke states of the spin ensemble initializing the superradiant Rabi oscillations (inset). Panels (d-f) show similar results as in the panels (a-c), except that the populations in the panel (d) refer to the  $T_1$ state spin levels, the lower inset of the panel (e) shows the zoom-in of the intra-resonator photon number for longer time, and the pentacene molecules are pumped by a $590$ nm laser pulse~\citep{HWu} with $2\times 10^3$ W power and $200$ $\mu$s duration, which in practice can be replaced by the Xenon flash lamp as in the experiments~\citep{HWu1}.  The parameters are taken from Ref.~\citep{Breeze} for the NV centers systems, and Refs.~\citep{Oxborrow, HWu1} for the pentacene molecules systems, see Tab. \ref{fig:tab} of the Appendix \ref{sec:Master-NV}. }
\end{figure*}

\section{Laser Power-Controlled Superradiant Rabi Oscillations~\label{sec:rabi}}

In the following, we apply the theories described above to study the dynamics of the NV centers/pentacene molecules-microwave resonator system. Before presenting the results, we stress that the optical pumping rate reflects better  the strength of the optical excitation while the laser power should be considered as a reference, because the relation between them might vary orders of magnitude in different experiments (see later discussion on Fig.~\ref{fig:steady-state}). For the NV centers and pentacene molecules system, we consider the illumination with $532$ nm laser~\citep{Breeze} and $590$ nm laser~\citep{HWu}, where the power up to  $20$ W might be achieved in the experiments~\citep{ASarkar}. If much larger power is needed, the Xenon lamp might be used instead~\citep{HWu1}. For high laser power, the temperature of solid-state might raise~\citep{CSzczuka}, which might lead to the draft of the spin transition frequency~\citep{VMAcosta}, the reduced spin coherence time~\citep{AJarmola}, and part of negatively NV centers might be converted into neutral NV~\citep{XDChen}. All these can introduce further complexity for the system response, but they can be principally incorporated into our models by introducing the  laser-power dependence of these parameters. However, these studies are beyond the scope of the current study, and will be examined in future.

\subsection{System with NV Centers}

Using the experimental parameters~\citep{Breeze}, we investigate the dynamics of the NV centers-microwave resonator system under the excitation of a pulsed laser with a power $2$ W and a duration $12$ ms [Fig. \ref{fig:rabi} (a-c)]. We study firstly the populations of the three ground-state spin levels [Fig. \ref{fig:rabi} (a)]. Note that the populations on the excited states are negligible and are thus not shown here. These spin levels are equally populated initially due to the spin-lattice relaxation. During the laser pumping, the population of the $m_s=0$ spin level increases to a constant value $0.48$, and the populations of $m_s=-1$ and $+1$ spin levels decrease to $0.28$ and $0.24$, respectively. After the laser pumping, the populations return gradually back to the thermal values. More importantly, the populations of the $m_s=0$ and $-1$ spin levels show small oscillations when the laser is switched on. Here, the population of the $m_s=-1$ spin level is slightly larger than that of the $m_s=+1$ spin level because the stimulated emission and absorption of photons tend to balance the populations of the $m_s=0$ and $-1$ spin levels. To reveal the related collective dynamics, we investigate also the equivalent dynamics of the spin ensemble in the Dicke state space [inset of Fig. \ref{fig:rabi} (b)], and find that the ensemble occupies initially the states with lowest symmetry at the left-most corner of Dicke state space, and then moves along the upper boundary of the Dicke triangle, and finally moves zigzag into the Dicke states with higher symmetry but slightly below the boundary. Note that here the Dicke state numbers are defined with respect to the $0\to -1$ spin transition, which couples resonantly to the microwave resonator.

The spin dynamics detailed above leads to the change of mean photon number inside the resonator shown in Fig. \ref{fig:rabi} (b). Before the laser excitation, the mean photon number is $662$ due to thermal excitation. When the laser is switched on, the mean photon number increases dramatically in about $2$ millisecond (not visible in the normal plot), and then oscillates with reduced amplitude on the order of $10^{9}$, forming firstly the superradiant Rabi oscillations, and converges gradually to a constant value around $5 \times 10^{8}$, leading to continuous-wave superradiant masing thereafter. When the laser is switched off, the photon number drops dramatically to the initial thermal value. To characterize the damped Rabi-oscillations, we calculate the frequency $\Omega$ of the oscillations from the period $1/\Omega$, and the duration $\tau$ to reach a factor of $1/e$ of the damped oscillation envelope. In Fig. \ref{fig:rabi} (c), we study the dependence of these parameters on the optical pumping rate (laser power). As the optical pumping rate (laser power) increases logarithmically from $2 \times 10^3$ Hz (2 W) to $1.1 \times 10^5$ Hz (100 W), the duration decreases gradually from $1.7$ ms to $0.04$ ms and the oscillation frequency increases gradually from $2$ kHz to $35$ kHz. Note that the time to reach the first peak of superradiant Rabi oscillations decreases also with the increased pumping rate (see Fig. \ref{fig:start} of the Appendix \ref{sec:extra}).  

To understand the superradiant Rabi oscillations, we notice that the spin ensemble occupies the states near the upper boundary of Dicke state space when the superradiant Rabi oscillations occur [inset of Fig. \ref{fig:rabi} (b)]. As explained in our previous study \citep{QWu2021}, these Dicke states can be well approximated as the states of upside-down quantized harmonic oscillators (according to the Holstein-Primakoff approximation~\citep{THolstein}), and the spin ensemble-resonator interaction can be described by the Hamiltonian $\hat{H} \approx \hbar \sum_{J} \sqrt{2J} g_s (\hat{a}^\dagger \hat{b}_J^\dagger + \hat{a} \hat{b}_J)$, where $g_s$ is the single spin-resonator coupling strength, $\hat{b}_J^\dagger, \hat{b}_J $ ($ \hat{a}^\dagger,\hat{a}$) are the creation and annihilation operators of the upside-down harmonic oscillators (resonator photons), respectively. This parametric Hamiltonian is responsible for the dramatic increase and decrease of the photon number, as opposite to the Jaynes-Cummings-type Hamiltonian as explored in our previous work~\citep{YZhang1}, and the collective coupling strength $\sqrt{2J}g_s$ determines the frequency of superradiant Rabi oscillations. The inset of Fig. \ref{fig:rabi} (c) shows that with stronger pumping, the spin ensemble occupies Dicke states with larger $J$ at the time just before the superradiant Rabi oscillations, which explains the increased frequency of the superradiant Rabi oscillations. 

\subsection{System with Pentacene Molecules}

Using the parameters compatible with the experiments ~\citep{JDBreeze,HWu1}, we obtain the results shown in Fig. \ref{fig:rabi} (d-f) under the excitation of a Xenon lamp flash with a power $2000$ W and duration $200$ $\mu$s. Such strong and long pumping radiation is considered in order to illustrate the intriguing dynamics over a longer time and the strong collective effects, see below. We analyze in Fig. \ref{fig:rabi} (d), the population dynamics of the spin sublevels of the  $T_1$ state. Note that the populations on the ground and excited states are negligible and thus are not shown here. We find that the population increases gradually from zero during the optical pumping, the population of the $X$ spin level increases faster than other spin levels, leading to the population inversion. After about $0.5$ ${\rm \mu s}$, the populations of the $X$ and $Z$ spin levels start to oscillate, and the oscillation amplitude decreases gradually. At about $3$ ${\rm \mu s}$, the populations reach quasi-steady-state values, and the population of the $Z$ spin level is larger than that of the $X$ spin level, indicating the disappearance of population inversion. Note that the population of the $Y$ spin level increases gradually, and saturates at about $0.29$. However, at longer time, the populations of all spin levels approach constant values, and importantly the population of the $X$ spin level becomes slightly larger than that of the $Z$ spin level, indicating a reappearance of the population inversion. After the switch off of the Xenon lamp, the populations decay gradually, and the population of the $Z$ spin level becomes larger than that of the $X$ spin level, i.e. the loss of the population inversion.

In the upper inset of Fig. \ref{fig:rabi} (e), we show the equivalent dynamics of the molecular spin ensemble in the Dicke state space. Here, the Dicke state numbers are defined with respect to the $X \to Z$ spin transition, which couples resonantly to the microwave resonator. The molecular spin ensemble is initially optically pumped to the Dicke state of higher symmetry and excitation ($J\approx M \approx 0.25N$ with $N\approx 10^{17}$) on the upper boundary, and then decays almost vertically to the lower boundary before climbing almost vertically to the Dicke states slightly below the upper boundary, which are caused by the resonator-mediated collective decay and absorption of the spin ensemble. Finally, the spin ensemble repeats the similar dynamics with however tendency toward the Dicke states on the lower boundary near the leftmost corner at short time, which are caused by the incoherent transitions introduced by the spin dephasing~\citep{YZhang4}, but eventually climbs to the states on the upper boundary. The evolution as observed is rather complex, and significantly different from the simple evolution of NV center ensemble [see the comparison of the insets of Fig.~\ref{fig:rabi}(b) and (e)].

The spin dynamics is accompanied by the change of mean photon number shown in Fig. \ref{fig:rabi} (e). Before the optical pumping, there are about $4211$ photons inside the resonator due to the thermal excitation. After the switch on of the Xenon lamp, the mean photon number increases rapidly in about 0.5 ${\rm \mu s}$ (not visible in normal plot), and then oscillates with reduced amplitude due to the collective dynamics of the spin ensemble in the Dicke state space [upper inset of Fig. \ref{fig:rabi} (e)]. Although the superradiant Rabi oscillations look similar to those of NV centers, they are actually caused by a different dynamics. At longer time, the mean photon number converges to a constant value around $6 \times 10^{13}$ [lower inset of Fig. \ref{fig:rabi} (d)], and we observe the continuous-wave superradiant masing, because of the recovery of the population inversion and the resulting stimulated emission process at long time. When the Xenon lamp is switched off, the mean photon number drops dramatically and reaches values below the thermal value (Fig. \ref{fig:rabi-short} of the Appendix \ref{sec:extra}
), which is the microwave mode cooling~\citep{HWu}, because of the absorption of microwave photons by the spin ensemble.  In Fig. \ref{fig:rabi-short} of the Appendix \ref{sec:extra}, we show that the continuous-wave superradiant masing does not occur if we adopt the values of the spin decay and dephasing, as applied in \citep{JDBreeze}.

We also study the superradiant Rabi oscillation frequency and duration as function of the optical pumping rates (Xenon lamp power) [Fig. \ref{fig:rabi} (f)]. We see that as the optical pumping increases, the duration decays from $9$ ${\rm \mu s}$ to  $1$ ${\rm \mu s}$ and the oscillation frequency increases from about $2.7$ MHz to $5.9$ MHz, which are qualitatively similar to those of the NV centers except for larger frequency and shorter duration. This dependence is observed because the molecular spin ensemble is pumped to the Dicke states with larger $J$ at the time before the oscillations [inset of Fig. \ref{fig:rabi} (f)], which initializes the faster superradiant Rabi oscillations due to the increased coupling with the microwave resonator. The results as revealed here can be also achieved with short but strong laser pulses, similar to those in the experiments~\citep{JDBreeze}. In Fig. \ref{fig:start} of the Appendix \ref{sec:extra}, we further analyze the delay time to reach the first peak of superradiant Rabi oscillations decreases with the increased pumping rate.

\begin{figure*}[!htp]
\begin{centering}
\includegraphics[scale=0.36]{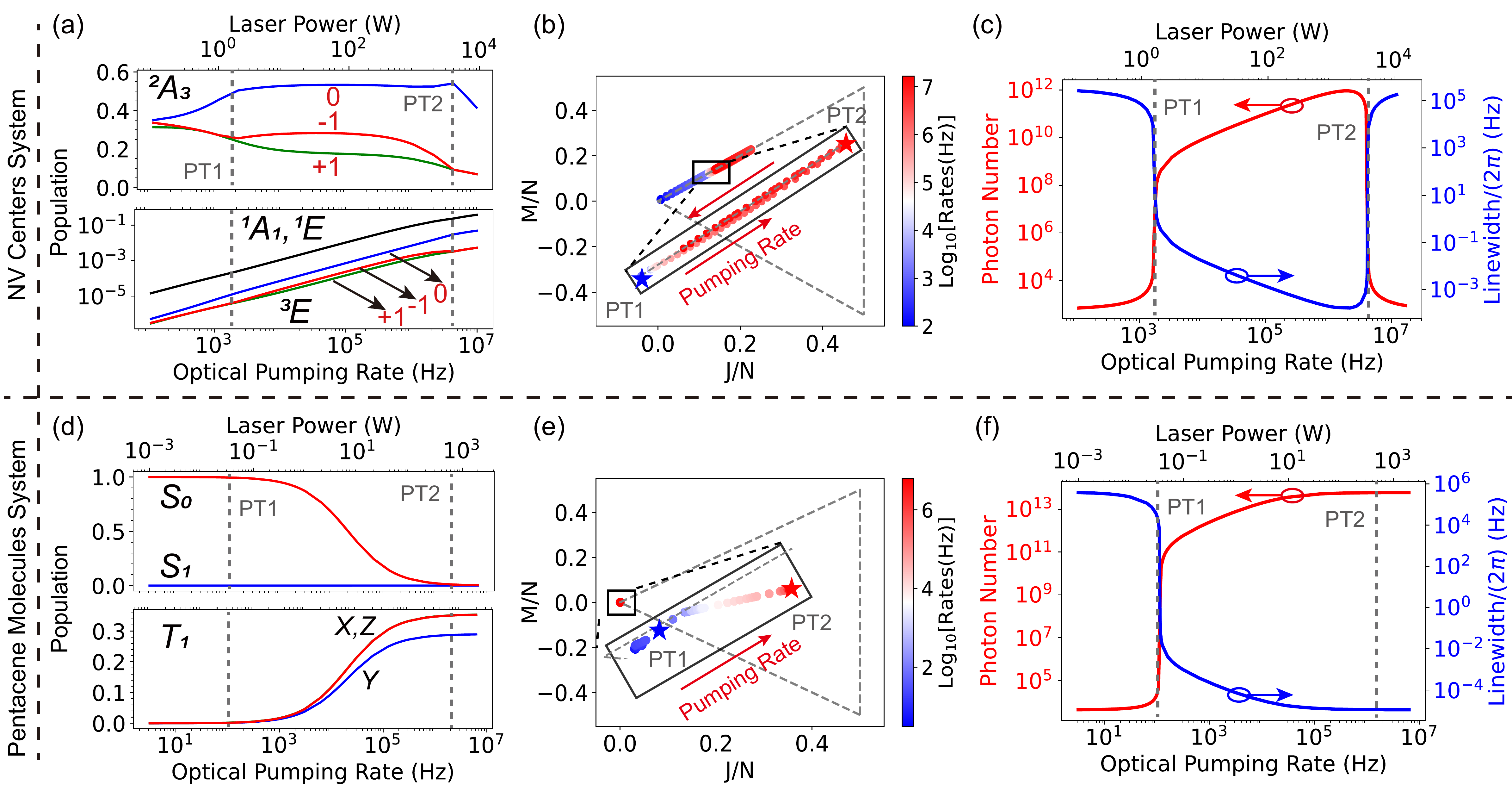}
\par\end{centering}
\caption{\label{fig:steady-state} Laser-power dependence of the steady-state of the systems with NV centers (a-c) and pentacene molecules (d-f) at room temperature. Panel (a,d) show the populations of all the electronic spin levels (as marked) as function of optical pumping rate (lower axes) or laser power  (upper axes). For the NV centers and pentacene molecules system, we consider the illumination with continuous-wave $532$ nm laser~\citep{WNg} and $590$ nm laser~\citep{HWu}, respectively, where the value up to $20$ W might be achieved in the experiments ~\citep{ASarkar}. In these and other panels, the dashed lines indicate the first and second optical pumping thresholds, denoted as PT1 and PT2. Panel (b,e) show the average Dicke state quantum numbers $J,M$ normalized to the number of NV centers/pentacene molecules $N$ for different optical pumping rates (color bar), where the blue and red stars mark the first and second pumping threshold, and the dashed lines indicate the boundaries of the Dicke state space. Panel (c,f) show the intra-resonator photon number (red solid line,left axis) and the radiation linewidth (blue solid line, right axis) as function of the optical pumping rate (lower axis) or the laser power (upper axis). Here, we consider that the relevant spin transitions of the NV centers and pentacene molecules are perfectly resonant with the microwave resonators, and will turn to  the off-resonant situations in Fig. \ref{fig:detuning}.  See Tab. \ref{fig:tab} of the Appendix \ref{sec:Master-NV} for other parameters.}
\end{figure*}

\section{Laser Power-dependence of continuous-wave Superradiant Masing~\label{sec:masing}}

In the previous section, we investigated the dynamics of the NV centers/pentacene molecules-microwave resonator systems under the laser/Xenon lamp pulse excitation, and concluded that both systems at the steady-state allow continuous-wave operation of a superradiant maser. In the following, we examine with Fig. \ref{fig:steady-state} the laser-power dependence of the electronic-spin levels populations (a,d), the states of the spin ensemble in the Dicke state space (b,e), and the intensity and linewidth of the superradiant masing (c,f). Since the superradiant masing intensity depends also on the out-coupling coefficient, we do not show it directly but display rather the intra-resonator photon number. Here, we assume that the relevant spin transitions of the NV centers and pentacene molecules are perfectly resonant with the microwave resonator. We return to the off-resonant situation in the next section. 

\subsection{System with NV centers}

\begin{figure}[!htp]
\begin{centering}
\includegraphics[scale=0.25]{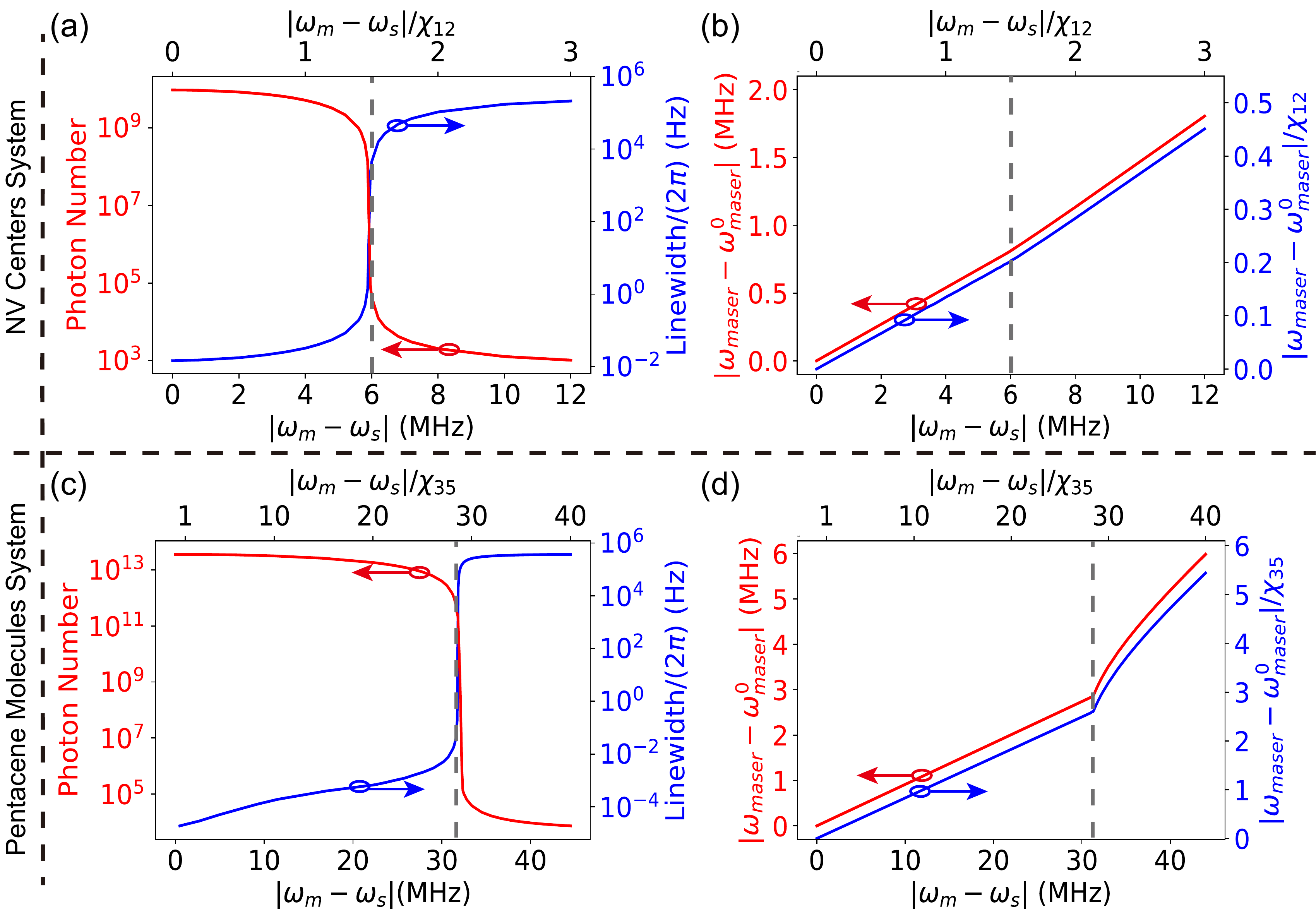}
\par\end{centering}
\caption{\label{fig:detuning} Influence of the frequency detuning on superradiant masing for systems with NV centers (a,b) and pentacene molecules (c,d) at room temperature for the continuous-wave $532$ nm~\citep{WNg} laser and $590$ nm laser ~\citep{HWu} excitation with realistic power $10$ W. Panel (a) and (c) show the intra-resonator photon number (left axis) and the linewidth (right axis) of the superradiant masing as a function of the detuning $|\omega_m-\omega_s|$ of the microwave resonator frequency $\omega_m$ and the spin transition frequency $\omega_s$ (lower axis). Panel (b) and (d) show the shift $|\omega_{\rm maser} - \omega^0_{\rm maser}|$ of the superradiant masing frequency $\omega_{\rm maser}$ relative to the one $\omega^0_{\rm maser} =\omega_m=\omega_s$ in the resonant case (left axis), indicating the cavity pulling effect. In all the panels, the upper axis indicates the ratio $|\omega_m-\omega_s|/\chi_{12},|\omega_m-\omega_s|/\chi_{35}$ of the frequency detuning to the dephasing rates of the spin transition $\chi_{12},\chi_{35}$. See Tab. \ref{fig:tab} of the Appendix \ref{sec:Master-NV} for other parameters.}
\end{figure}

Fig.~\ref{fig:steady-state} (a) shows that for the NV centers under extremely weak laser excitation, the populations are almost equally distributed between the ground state spin levels (upper part), and the populations of the excited states are negligible (lower part). As the optical pumping rate (laser power) increases to about $1.6$ kHz ($1.5$ W), the population of the ground state $m_s=0$ ($m_s=\pm 1$) spin level increases (decreases) gradually to $0.49$ ($0.25$), and the populations of the excited states increase also but are still orders of magnitude smaller. When the optical pumping rate  (laser power) increases further to $4$ MHz ($4\times 10^3$ W), the population of the ground state $m_s=0$ spin level remains at the finite value, that of the ground state $m_s=-1$ spin level remains at the finite value and then decreases, and that of the ground state $m_s=+1$ spin level decreases gradually. At the same time, the populations of the excited states increase gradually, and in particular that of the singlet excited states $^1A_1, ^1E$ approach $0.2$ and thus become comparable with the populations of the ground state spin levels. If we further increase the optical pumping rate (laser power), the populations of the ground state spin levels decrease dramatically, while those of the excited levels increase gradually. From the above results, we find that optical pumping rates (laser powers) at $1.6$ kHz ($1.5$ W) and $4$ MHz ($4\times 10^3$ W) are threshold values, where qualitative changes occur to the spin ensemble. By comparing the first threshold with the value $0.15$ W as reported in the experiments~\citep{Breeze}, the conversion from the laser power to the optical pumping rate, as considered here, might be underestimated by one order of magnitude. If taking this correction into account, all the laser powers as given from Fig.~\ref{fig:rabi} to \ref{fig:detuning} should be reduced by one order of magnitude for the systems with NV centers.

In Fig. \ref{fig:steady-state}(b), we show the equivalent evolution for the NV spin ensemble in the Dicke state space for different optical pumping rates. Note that here the Dicke state numbers are defined with respect to the $0\to -1$ spin transition. We see that as the optical pumping rate increases gradually to the first threshold value (blue star), the spin ensemble starts from the Dicke state with lowest symmetry at the leftmost corner, and climbs upwards along the upper boundary to the Dicke state with $J\approx M\approx 0.12 N$. When the optical pumping rate increases further to the second threshold (red star), the spin ensemble moves further to the Dicke state with $J\approx M\approx 0.23 N$. For much larger optical pumping rate, the spin ensemble decreases along the upper boundary. Here, the spin ensemble does not reach the uppermost Dicke states because the population transfers to the higher excited states, as identified in our previous work~\citep{YZhang1}, which is in contrast to the predictions based on the effective two-level models for the NV center spins~\citep{QWu2021}. By analyzing carefully the results between the two thresholds, we find that the spin ensemble does not fully follow the upper boundary but occupies states slightly below the boundary. Considering the larger number of NV centers involved $N\approx 4 \times 10^{13}$, the slight deviation here means $10^{11}$ of Dicke states, and thus the superradiance of the NV spin ensemble might also contribute to the coherent masing.

The evolution of the NV spin ensemble leads to the change of the superradiant masing  as shown in Fig. \ref{fig:steady-state}(c). We see that as the optical pumping rate (laser power) increases slowly to and over the first threshold, the intra-resonator photon number increases gradually from the thermal value around $661$ to about $10^4$ and jumps dramatically to $10^8$, while the radiation linewidth reduces steadily from about $260$ kHz to about $10$ kHz, and then decays abruptly to $0.3$ Hz. As the optical pumping rate (laser power) increases further and over the second threshold, the photon number increases slowly again to about $10^{12}$ and then drops dramatically below $10^4$, while the radiation linewidth reduces steadily again to about $10^{-4}$ Hz and then jumps dramatically over $10^4$ Hz. By associating this result with Fig. \ref{fig:steady-state}(b), we might conclude that the superradiant masing with minimal linewidth is achieved for the spin ensemble at the Dicke states with larger $J$ close to the state at the second threshold (see the red star). Here, the predicted minimal linewidth and the corresponding Dicke states are in accord with the results in our previous study~\citep{QWu2021}.

\subsection{System with Pentacene Molecules}

Fig.~\ref{fig:steady-state} (d) shows that for the pentacene molecules system under weak laser excitation, the population is mostly on the electronic ground state (upper part), and the populations of the spin levels $X,Y,Z$ of the lowest triplet state $T_1$ are negligible (lower part). As the optical pumping rate (laser power) increases from about $10^2$ Hz ($3.5\times 10^{-2}$ W), the population of the singlet ground state decreases gradually, while the populations of the triplet spin levels increase steadily. When the optical pumping rate (laser power) approaches $1.8\times 10^6$ Hz ($600$ W), the population of the ground state approaches zero, while those of the spin levels approach their saturated values. Note that the population of the excited state is always near zero for the laser pumping considered here due to the ultra-fast inter-system crossing from this state to the $T_1$ state. In addition, the populations of the $X$ and $Z$ spin levels are almost identical, and are always larger than that of the $Y$ spin level. A more careful analysis indicates the population  of the $X$ spin level is always slightly larger than that of the $Z$ spin level, forming the population inversion between these levels. Here, we also identify two thresholds, where the populations change qualitatively. This is similar to the NV centers system except that the population of the spin levels will not decrease for the laser excitation exceeding the second threshold.

Fig. \ref{fig:steady-state} (e) shows the equivalent evolution of the molecular spin ensemble in the Dicke state space for increasing optical pumping rate. As before, the Dicke state numbers are defined with respect to the $X\to Z$ transition. We find that the spin ensemble always occupies the states of lower symmetry near the leftmost corner of the Dicke state space. By examining the results more carefully [inset of Fig. \ref{fig:steady-state} (e)], we identify that the spin ensemble starts initially from the Dicke states below the upper boundary, and moves toward this boundary as the optical pumping rate approaches the first threshold value (blue star), and then away from the boundary to explore the states inside the Dicke state space, and finally stacks at the specific states as the optical pumping rate approaches the second threshold (red star). As before, this departure indicates that the superradiance also contributes to the coherent radiation. These results are completely different from those for the system with NV centers. 

The spin dynamics revealed by our analysis, leads to the laser-power dependence of the superradiant masing intensity and linewidth as shown in Fig. \ref{fig:steady-state} (f). We see that as the optical pumping rate (laser power) approaches the first threshold, the intra-resonator photon number increases slowly from about $4\times 10^3$ to about $10^5$, and then jumps dramatically to about $5 \times 10^{10}$. At the same time, the radiation linewidth reduces from about $400$ kHz to $5$ kHz, and then drops abruptly to about $10$ mHz. As the optical pumping rate (laser power) increases further to and over the second threshold, the photon number increases steadily and approaches a saturated value around $6\times 10^{13}$, while the radiation linewidth reduces steadily to the constant value $0.01$ mHz. We have also checked that the saturated behavior does not change for much larger laser power, e.g. $10^3$ W (not shown). In comparison to Fig. \ref{fig:steady-state}(c), we see that the pentacene molecules system could achieve superradiant masing with stronger intensity and narrower linewidth for orders of magnitude smaller optical pumping rate (laser power). Here, the lower laser intensity required might be attributed to the fact the spin levels are on the triplet excited state of the pentacene molecule, and are not so strongly affected by the spin-lattice relaxation.

\section{Influence of Spins-Resonator Frequency Detuning on Superradiant Masing~\label{sec:pulling}}

In the previous section, we have investigated the superradiant masing on the ideal situations where the relevant spin transitions are perfectly resonant with the microwave resonators. However, in realistic systems, due to magnetic field variations and resonator thermal fluctuations, the frequency of the spin transitions and the resonators may drift or show fluctuations. Thus, it is necessary to analyze the influence of the spin transition-microwave resonator frequency detuning on the superradiant masing. 

In Fig.~\ref{fig:detuning}, we study this influence on the NV centers system (a,b) and the pentacene molecules system (c,d) under the continuous-wave laser pumping with the realistic $10$ W power, which can be achieved in the experiments. Fig. \ref{fig:detuning}(a) shows that as the frequency detuning overcomes $1.5$ times the dephasing rate of the spin transition $1.5\chi_{12}\approx2\pi\times0.96$ MHz (lower and upper axis), the intra-resonator photon number (left axis) reduces firstly moderately, and then drops dramatically by several orders of magnitude, and finally decreases steadily to the thermal value. At the same time, the radiation linewidth (right axis) increases slowly within millihertz, then climbs abruptly by orders of magnitude, and finally raises steadily above kilohertz. These results indicate that the superradiant masing with large intensity and ultra-narrow linewidth can be sustained as far as the frequency detuning is within the dephasing rate of the spin transition. 

Furthermore, Fig. \ref{fig:detuning}(b) shows that the shift of the superradiant masing frequency from that in the ideal situation increases linearly with the frequency detuning. From this result, we can estimate the so-called cavity pulling factor, as the ratio of the shift to the frequency detuning, to be around $0.13$, which is smaller than unity. It was expected that this factor should be orders of magnitude smaller than unity for the steady-state superradiance, which relies only on the coherence stored in the gain medium ~\citep{DMeiser2009,JDBohnet2012}, but it should be close to unity for the lasing or masing, which depends only on the coherence in the photons. Thus, the coherent radiation achieved here can be classified as superradiant masing, where the coherence in both the gain medium and the photons contribute~\citep{KDebnath2018}. However, since the cavity pulling factor is not orders of magnitude smaller than unity, we expect that the field coherence dominates the radiation coherence. 

We study now the pentacene molecules system. Fig. \ref{fig:detuning}(c) shows that as the frequency detuning increases, the intra-resonator photon number and the radiation linewidth behave similar as Fig. \ref{fig:detuning}(a) for the NV centers system, except that the superradiant masing can be sustained for the frequency detuning below around $30$ MHz, which is about five times larger. Importantly, we find that this detuning range is not determined by the dephasing of the spin transition $\chi_{35}$ but roughly by the collective spin-microwave mode coupling $\sqrt{2J}g$. Furthermore, Fig. \ref{fig:detuning}(d) shows that the shift of the superradiant masing frequency with respect to that in the resonant case behaves similar as Fig. \ref{fig:detuning}(b) except that it increases dramatically for the larger frequency detuning. From the results, we estimate the cavity pulling factor as $0.09$, which is slightly smaller than the value for the NV center systems. This suggests that in the pentacene molecule systems, the coherence stored in the gain medium or the superradiance contributes slightly more to the coherent radiation.

\section{Discussions and Conclusions \label{sec:conclusions}}

In summary, we propose that solid-state spins, e.g. pentacene molecules and NV centers, coupled to the microwave resonators with high quality factor can be used to explore steady-state superradiance and superradiant masing at room temperature. To verify our proposal, we have developed the quantum master equations to describe the complex systems involving multiple levels, multiple processes and trillions of spins, and employ the second-order mean-field approach to analyze the laser-power dependence of the system dynamics, the spin ensemble Dicke states, and the steady-state properties, e.g. the intra-resonator photon number and the radiation linewidth.  

Our calculations predict that the superradiant Rabi oscillations occur prior to the continuous-wave superradiant masing, and are caused by the collective transitions among different Dicke states for the systems with NV centers and pentacene molecules. In addition, the frequency and duration of the oscillations can be actively controlled by the laser power via pumping the spin ensemble to different regions of  Dicke states, which initialize the superradiant Rabi oscillations. The continuous-wave superradiant masing occurs for moderate laser pumping, and  can have a linewidth well below millihertz. Furthermore, the superradiant masing is caused by the states inside the Dicke state space, and the calculated cavity pulling factor is about one order of magnitude smaller than unity, which both indicate the contribution of superradiance besides the dominated contribution from the stimulated emission to the coherent radiation. In comparison to the optical lattice systems, our system suffers from relatively large dephasing, and thus the dynamics and steady-state are mainly dominated by the states near the upper boundary of the Dicke state space. Thus, further theoretical study should be carried out to understand how the  increasing dephasing might affect the steady-state superradiance and the superradiant lasing/masing, which is however out the scope of the current study.

Our work sheds light into the transient and steady-state superradiant masing, and it may guide further explorations on the aspect of ultra-narrow radiation, which might find applications in deep-space communications, radio astronomy and high-precision metrology. Furthermore, the mean-field theory and the Dicke state presentation, as advocated in this study, can be applied also to investigate similar phenomena in the systems with other solid-state spins, such as silicon vacancy centers in silicon carbide~\citep{HKraus,MFischer} and boron vacancy centers in hexagonal boron nitride~\citep{AGottscholl2020,AGottscholl2021}.

\begin{acknowledgments}
This work was supported by the National Natural Science Foundation of China project No. 12004344, 62027816, and Henan Center for Outstanding Overseas Scientists project No. GZS201903, as well as the Danish National Research Foundation through the Center of Excellence for Complex Quantum Systems (Grant agreement No. DNRF156), and the European Union's Horizon 2020 Research and Innovation Programme under the Marie Sklodowska-Curie program (No. 754513). Yuan Zhang convinced the idea and theory, Qilong Wu implemented the numerical calculations, and they contributed equally to this work. All the authors contributed to the analyses of the results and to the writing of the manuscript. 
\end{acknowledgments}

\appendix 
\renewcommand\thefigure{A\arabic{figure}}
\renewcommand\thetable{A\arabic{table}}
\renewcommand{\bibnumfmt}[1]{[S#1]}
\setcounter{figure}{0}  

\section{Quantum Master Equations and Julia Codes for Solid-state spins-Microwave Resonators Systems \label{sec:Master-NV}}

In this Appendix, we present the quantum master equations for the dynamics of the solid-state spin-microwave resonator systems, and the Julia codes to solve these equations with the mean-field approach.

\begin{figure}[!htp]
\begin{centering}
\includegraphics[scale=0.5]{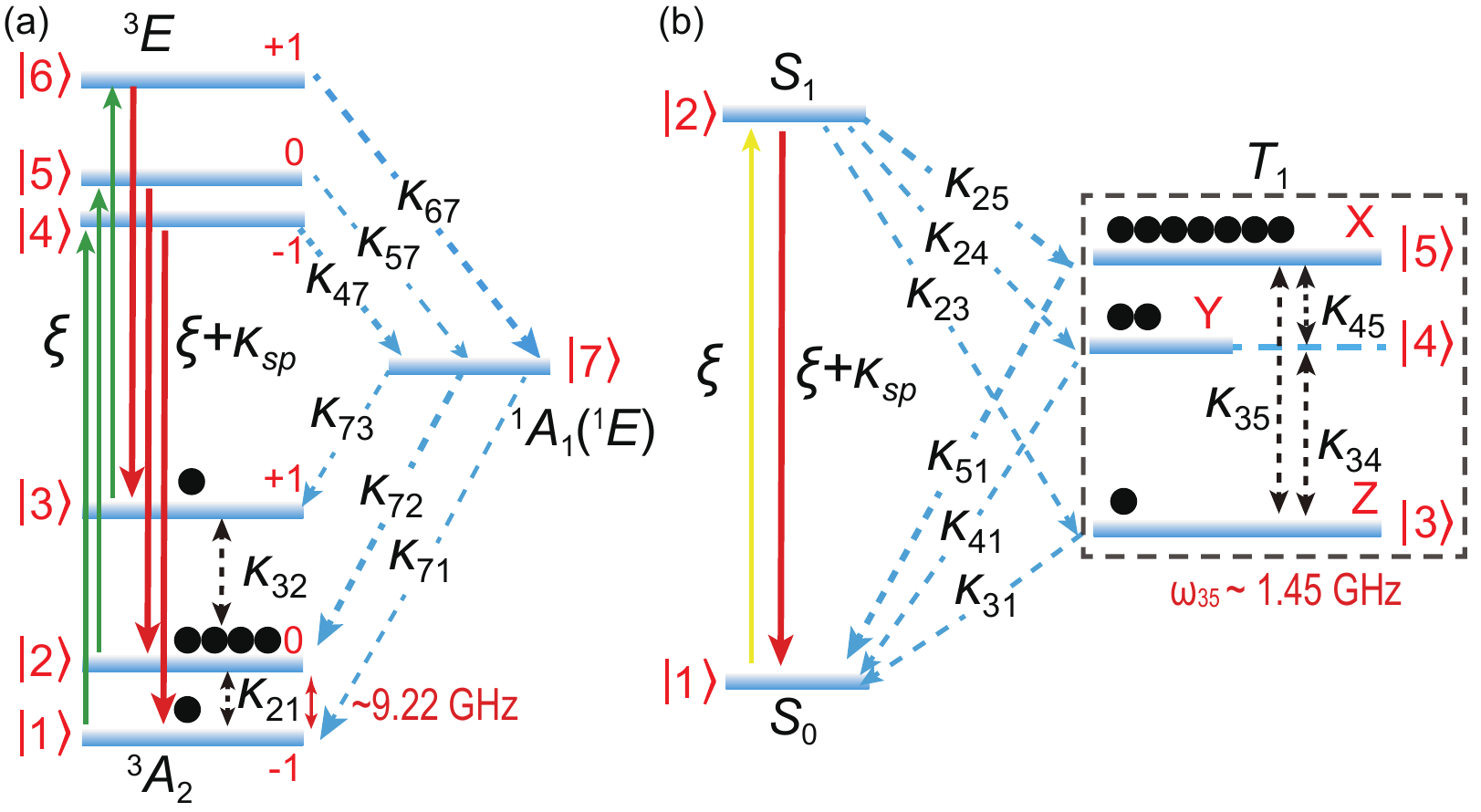}
\par\end{centering}
\caption{\label{fig:com} Labelings of the energy levels and the transition rates among them for the NV centers in diamond (a) and the pentacene molecules (b). $\xi,k_{sp}$ represent the optical pumping and the spontaneous emission, respectively, while other symbols $k_{ij}$ indicate the non-radiative decay processes. $\chi_{ij}$ label the spin dephasing between the spin levels of same electronic level (not shown). The labelings are used in the Julia codes to define and solve the quantum master equation with the mean-field approach.}
\end{figure}

\begin{table*}[!htp]
\begin{centering}
\includegraphics[scale=0.5]{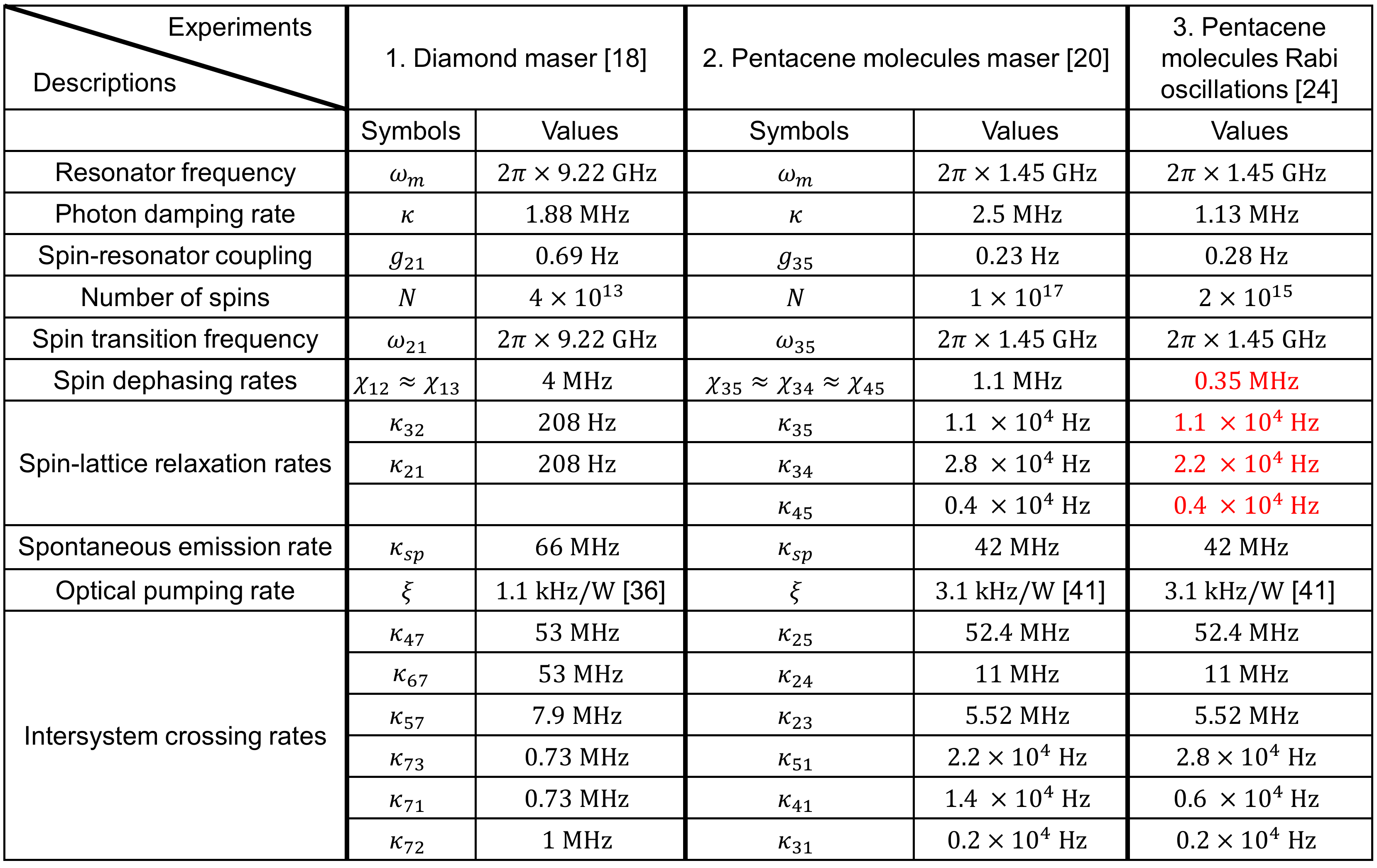}
\par\end{centering}
\caption{\label{fig:tab} Summary of the parameters for the NV centers-microwave resonator system (second and third columns), taken from the diamond maser experiment~\citep{Breeze}, and for the pentacene molecules-microwave resonator system (fourth and fifth columns), taken from the quasi-continuous-wave masing experiment~\citep{HWu1}. The last column indicates the parameters for the latter system from the Rabi oscillations experiment~\citep{JDBreeze}, where the spin decay and dephasing rate are not representative of more recent experiments (marked with red fonts). The optical pumping rate $\xi$ is estimated with the parameters for the diamond and pentacene sample as used in the experiments~\citep{WNg,HWu}, respectively. 
}
\end{table*}

\subsection{Labeling of Levels and Transition Rates \label{sec:para}}

In Fig. 1 (b) of the main text, we have shown the energy levels and the transitions among them for the NV centers. To construct the quantum master equation and apply the mean-field approach, we label these levels and transition rates as shown in Fig. \ref{fig:com} (a). We label the three spin levels with projections $m_{s}=-1,0,1$ of the triplet electronic ground state $^{3}A_{2}$ as levels $i=1,2,3$, and of the triplet electronic excited state $^{3}E$ as levels $i=4,5,6$, and we represent the two singlet excited states $^{1}E, {^{1}A_1}$ as one representative level $i=7$ \citep{WNg,YZhang1}. We consider the spin-preserving optical pumping and spontaneous emission rates $\xi,k_{sp}$, and the inter-system crossing rates $k_{i7},k_{7j}$ from the spin levels $i=4,5,6$, and to the spin levels $j=1,2,3$. We consider also the spin-lattice relaxation rates $k_{32},k_{21}$ between the $m_{s}=0$ spin level and the $m_{s}=\pm1$ spin levels, and the corresponding dephasing rates $\chi_{32},\chi_{21}$. In addition, we consider a microwave resonator with a frequency $\omega_m$ and a photon damping rate $\kappa$, which couples resonantly to the $2 \leftrightarrow 1$ transition of the NV centers with a strength $g_{21}$. In our simulations, we take the parameters from the experiment~\citep{Breeze}, and summarize them in the second and third columns of Tab. \ref{fig:tab}.

The optical pumping rate $\xi$ can be calculated with the expression 
\begin{equation} 
\xi=\frac{\lambda_{P}\sigma_{\lambda_{P}}}{hcA_{p}l\alpha}\left(1-e^{-l\alpha}\right)\left(1-R\right)P. \label{eq:pumping}
\end{equation} 
For the NV centers system \cite{WNg}, $\lambda_{P}=532$ nm is the wavelength of the pumping laser, $\sigma_{\lambda_{P}}=3.1\times10^{-21}\mathrm{m}^{2}, \alpha=2.3\times10^{3}$ $\mathrm{m}^{-1}$ are the absorption cross-section and coefficient of the NV centers at $\lambda_{P}$. $h$ and $c$ represent Planck's constant and the speed of light, respectively. $A_{p}=1.76\times10^{-6}$  $\mathrm{m^{2}}$ is the cross-sectional area of the pump beam incident on the sample. $l=1.5$ mm is the thickness of the diamond crystal. $R=\left|\frac{n_{1}-n_{2}}{n_{1}+n_{2}}\right|^2$ is the Fresnel reflection coefficient, $n_{1}=1$ is the refractive index of air and $n_2 =2.42$ is the refractive index of diamond, $P$ is the pumping laser power.

In Fig. 2 (d) of the main text, we have shown the energy levels and the transitions among them for the pentacene molecules. We label these levels and the rates of transitions in Fig. \ref{fig:com} (b). We label the singlet electronic ground $S_{0}$ and the excited state $S_{1}$ as levels $1,2$, and the spin states $Z,Y,X$ of the lowest triplet state $T_{1}$ as levels $3,4,5$. We consider the optical pumping and the spontaneous emission rates $\xi,k_{sp}$, the inter-system crossing rates $k_{2i},k_{j1}$ to and from the spin levels $i,j=3,4,5$ of the $T_1$ state, and the spin-lattice relaxation rates $k_{35},k_{45},k_{34}$ as well as the dephasing rates $\chi_{35},\chi_{45},\chi_{34}$. In addition, we consider a microwave resonator with a frequency $\omega_m$ and a photon damping rate $\kappa$, which couples resonantly to the $5 \leftrightarrow 3$ transition of the pentacene molecules with the strength $g_{35}$. The value of the aforementioned parameters are taken from the quasi-maser experiment~\citep{HWu1}, see the third and fourth columns of Tab.  \ref{fig:tab}. The optical pumping rate $\xi$ of the pentacene molecules can be calculated by the simplified version of Eq. (\ref{eq:pumping}):
$
\xi \approx \frac{\lambda_{P}\sigma_{\lambda_{P}}}{hcA_{p}}P 
$ 
with $\lambda_{P}=590$ nm, $\sigma_{\lambda_P}\approx 2\times10^{-21}\mathrm{m}^{2}, A_{p}=1.9\times10^{-6} \mathrm{m^{2}}$~\citep{HWu} .

\subsection{Quantum Master Equations} \label{sec:multilevelJC}

To describe the dynamics of the NV centers-microwave resonator system, as shown in Fig. \ref{fig:com} (a), we introduce the following quantum master equation for the reduced density operator $\hat{\rho}$:
\begin{align}
 & \partial_{t}\hat{\rho}=-\frac{i}{\hbar}\left[\hat{H}_{NV}+\hat{H}_{m}+\hat{H}_{m-NV},\hat{\rho}\right]\nonumber \\
 & -\xi\sum_{k}\left(\mathcal{D}\left[\hat{\sigma}_{k}^{41}\right]\hat{\rho}+\mathcal{D}\left[\hat{\sigma}_{k}^{52}\right]\hat{\rho}+\mathcal{D}\left[\hat{\sigma}_{k}^{63}\right]\hat{\rho}\right)\nonumber \\
 & -\left(\xi+k_{sp}\right)\sum_{k}\left(\mathcal{D}\left[\hat{\sigma}_{k}^{14}\right]\hat{\rho}+\mathcal{D}\left[\hat{\sigma}_{k}^{25}\right]\hat{\rho}+\mathcal{D}\left[\hat{\sigma}_{k}^{36}\right]\hat{\rho}\right)\nonumber \\
 & -\sum_{k} (\sum_{i=4,5,6}k_{i7}\mathcal{D}\left[\hat{\sigma}_{k}^{7i}\right]\hat{\rho} + \sum_{i=1,2,3}k_{7i}\mathcal{D}\left[\hat{\sigma}_{k}^{i7}\right]\hat{\rho} )   \nonumber \\
 & -\sum_{k}\sum_{i=1,3}\left( k_{i2}\mathcal{D}\left[\hat{\sigma}_{k}^{2i}\right]\hat{\rho}+k_{2i}\mathcal{D}\left[\hat{\sigma}_{k}^{i2}\right]\hat{\rho} \right) \nonumber \\
& -\sum_{k}\sum_{i=1,3}\left(\frac{1}{2} \chi_{i}\mathcal{D}\left[\hat{\sigma}_{k}^{ii}-\hat{\sigma}_{k}^{22}\right]\hat{\rho}\right) \nonumber \\
 & -\kappa\left[\left(n_{m}^{th}+1\right)\mathcal{D}\left[\hat{a}\right]\hat{\rho}+n_{m}^{th}\mathcal{D}\left[\hat{a}^{\dagger}\right]\hat{\rho}\right].\label{eq:me-nv}
\end{align}
The Hamiltonian $\hat{H}_{NV}=\hbar \omega_{21}\sum_{k=1}^{N}\hat{\sigma}_{k}^{21}\hat{\sigma}_{k}^{12}$ describes the spin transition $1 \leftrightarrow 2$ of the NV centers, which couples resonantly to the microwave resonator, and is determined by the frequency $\omega_{21}$ and the raising $\hat{\sigma}_{k}^{21}$ and lowering operator $\hat{\sigma}_{k}^{12}$. Here, the label $k,N$ indicate the individual NV center and the number of NV centers, respectively. The operators $\hat{\sigma}_{k}^{ij}=\left|i_{k}\right\rangle \left\langle j_{k}\right|$ represent projection operators for $i=j$ and transition operators for $i\neq j$. The Hamiltonian $\hat{H}_{m}=\hbar\omega_{m}\hat{a}^{\dagger}\hat{a}$ describes the microwave resonator mode with the frequency $\omega_{m}$, the photon creation $\hat{a}^{\dagger}$ and annihilation operator $\hat{a}$. The Hamiltonian $\hat{H}_{NV-m}=\hbar g_{21}\sum_{k}\left(\hat{a}^{\dagger}\hat{\sigma}_{k}^{12}+\hat{\sigma}_{k}^{21}\hat{a}\right)$ describes the interaction between the NV center spins and the microwave resonator with the strength $g_{21}$.

The second and third line of Eq. (\ref{eq:me-nv}) describe the optical excitation and the spontaneous emission of the NV centers with the rates $\xi, k_{sp}$. Here, we introduce the Lindblad superoperator $\mathcal{D}\left[\hat{o}\right]\hat{\rho}=\frac{1}{2}\left\{ \hat{o}^{\dagger}\hat{o},\hat{\rho}\right\} -\hat{o}\hat{\rho}\hat{o}^{\dagger}$ for any operator $\hat{o}$. The fourth line describes the non-radiative decay to and from the (representative) singlet excited state through the inter-system crossing with the rates $k_{i7},k_{7j}$ (for $i=4,5,6$ and $j=1,2,3$), respectively. The fifth and sixth line describe the spin-lattice relaxation rates $k_{32} \approx k_{23},k_{21} \approx k_{12}$ between the spin levels on the ground state, and the spin dephasing with the rates $\chi_{13},\chi_{12}$, incorporating both the dephasing due to the interactions with the spin bath and the decoherence due to fluctuations of spin transition frequencies. The seventh line describes the stimulated, and spontaneous emission (former term) and the stimulated absorption of thermal photons (latter term) by the microwave resonator with the rate $\kappa$ and the thermal equilibrium photon number $n_{m}^{th}=\left[e^{\hbar\omega_{m}/k_{B}T}-1\right]^{-1}$ at temperature $T$ (with the Boltzmann constant $k_B$).

To describe the dynamics of the pentacene molecules-microwave resonator system, as shown in Fig. \ref{fig:com} (b), we introduce the following quantum master equation for the reduced density operator $\hat{\rho}$:  
\begin{align}
 & \partial_{t}\hat{\rho}=-\frac{i}{\hbar}\left[\hat{H}_{PM}+\hat{H}_{m}+\hat{H}_{m-PM},\hat{\rho}\right]\nonumber \\
 & -\xi\sum_{k}\mathcal{D}\left[\hat{\sigma}_{k}^{21}\right]\hat{\rho}\nonumber  -\left(\xi+k_{sp}\right)\sum_{k}\mathcal{D}\left[\hat{\sigma}_{k}^{12}\right]\hat{\rho}\nonumber \\
 & -\sum_{k} (\sum_{i=3,4,5}k_{2i}\mathcal{D}\left[\hat{\sigma}_{k}^{i2}\right]\hat{\rho} + \sum_{i=3,4,5}k_{i1}\mathcal{D}\left[\hat{\sigma}_{k}^{1i}\right]\hat{\rho}) \nonumber \\
 & -\sum_{k}\sum_{i,j=3,4,5;i \neq j}\left( k_{ij}\mathcal{D}\left[\hat{\sigma}_{k}^{ji}\right]\hat{\rho}\right ) \nonumber \\
 & -\sum_{k}\sum_{i,j=3,4,5;i \neq j}\left( \frac{1}{2} \chi_{ij}\mathcal{D}\left[\hat{\sigma}_{k}^{ii}-\hat{\sigma}_{k}^{jj}\right]\hat{\rho}\right ) \nonumber \\
 & -\kappa\left[\left(n_{m}^{th}+1\right)\mathcal{D}\left[\hat{a}\right]\hat{\rho}+n_{m}^{th}\mathcal{D}\left[\hat{a}^{\dagger}\right]\hat{\rho}\right].\label{eq:me-pm}
\end{align}
The Hamiltonian $\hat{H}_{PM}=\hbar\omega_{35}\sum_{k=1}^{N}\hat{\sigma}_{k}^{35}\hat{\sigma}_{k}^{53}$ describes the levels of the pentacene molecules with transition frequency $\omega_{35}$, which couple resonantly to the microwave resonator. The Hamiltonian $\hat{H}_{m}=\hbar\omega_{m}\hat{a}^{\dagger}\hat{a}$ of the microwave resonator has been introduced before. The Hamiltonian  $\hat{H}_{m-PM}=\hbar\sum_{k}g_{35}\left(\hat{\sigma}_{k}^{53}\hat{a}+\hat{a}^{\dagger}\hat{\sigma}_{k}^{35}\right)$ describes the interaction between the pentacene molecules and the microwave resonator with the coupling strength $g_{35}$.

In Eq. \eqref{eq:me-pm}, the second line describes the optical pumping and the spontaneous emission with the rates $\xi,k_{sp}$. The third line  contains the inter-system crossing with the rates $k_{2i},k_{i1}$ ($i=3,4,5$) to and from the triplet ground state. The fourth line  describes the spin-lattice relaxation 
with the rates $k_{34} \approx k_{43} ,k_{35} \approx k_{53},k_{45} \approx k_{54}$ between different spin levels. The fifth line describes the dephasing process with the rates $\chi_{35}, \chi_{45}, \chi_{34}$. The last line describes the thermal emission and excitation of the microwave resonator as in Eq. \eqref{eq:me-nv}.

\subsection{Julia Codes to Solve Quantum Master Equations with Mean-field Approach \label{sec:code-nv}}

\begin{figure*}[!htp]
\begin{centering}
\includegraphics[scale=0.32]{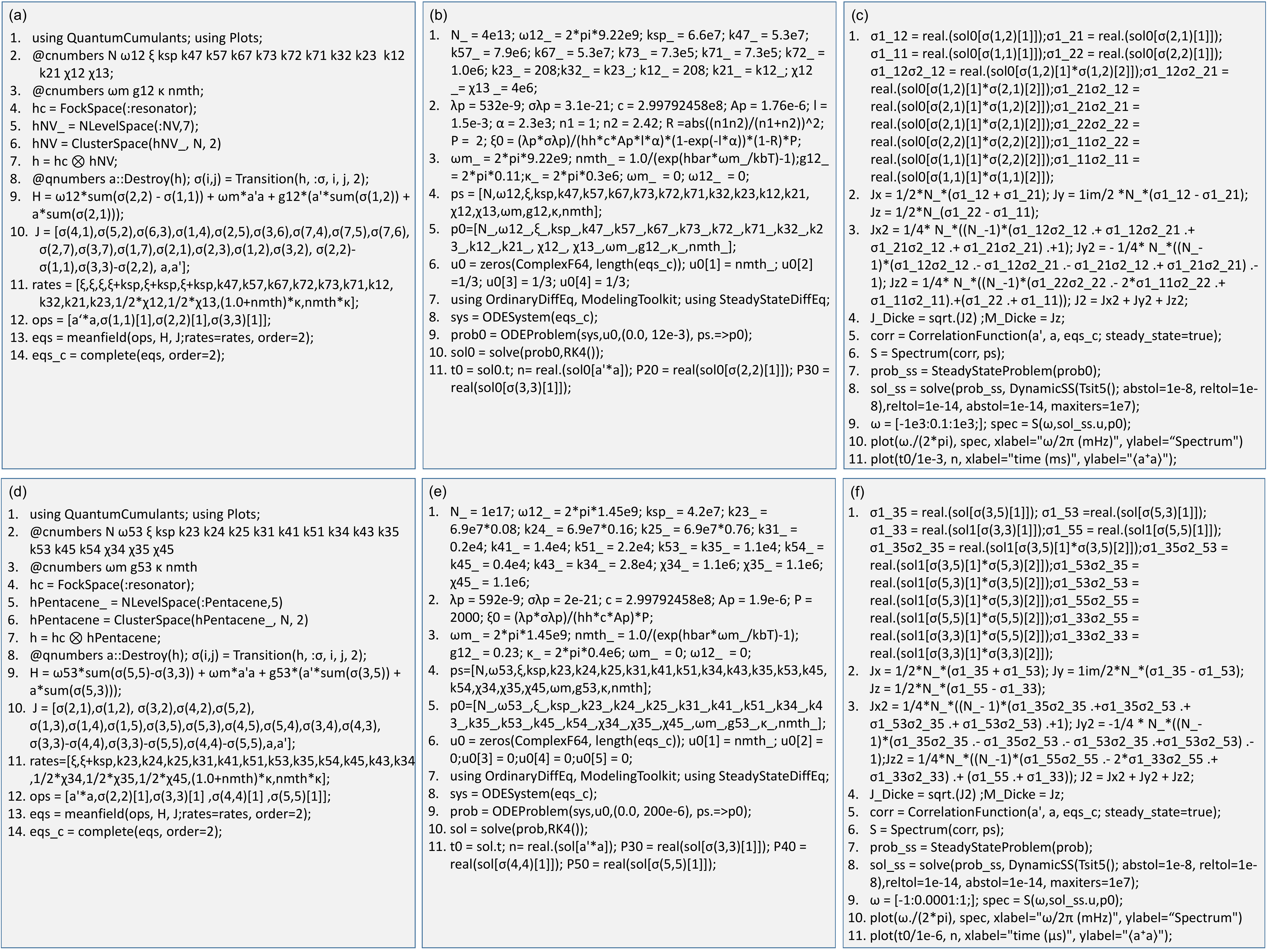}
\par\end{centering}
\caption{\label{fig:code-NV} Julia codes to derive the mean-field equations (a,d) and solve those equations (b,e), and compute the average of Dicke state quantum numbers and the radiation spectrum (c,f),  for the NV centers system (a-c) and the pentacene molecules system (d-f), respectively.}
\end{figure*}

As mentioned in the main text, we can not apply the standard density matrix technique to solve the master equations (\ref{eq:me-nv}) and (\ref{eq:me-pm}) due to the huge number of matrix elements involved, and have to rely on the mean-field approach. In this approach, we derive the equation $\partial_{t}\left\langle \hat{o}\right\rangle =\mathrm{tr}\left\{ \hat{o}\partial_{t}\hat{\rho}\right\} $ for the expectation values $\left\langle \hat{o}\right\rangle =\mathrm{tr}\left\{ \hat{o}\hat{\rho}\right\} $ of any operator $\hat{o}$, truncate the hierarchy of the coupled equations by using cumulant expansion, and reduce further the number of independent equations by assuming all NV centers or pentacene molecules identical. In practice, we use the QuantumCumulant.jl package \citep{DPl} to  derive and solve the mean-field equations, and we describe the Julia codes developed for this purpose in the following.

Fig. \ref{fig:code-NV} (a-c) show the Julia codes to derive the mean-field equations from the quantum master equation \eqref{eq:me-nv} for the NV centers system, the codes to convert the symbolic equations to numerical ones and then solve them numerically, respectively. Fig. \ref{fig:code-NV} (d-f) show the Julia codes for the pentacene molecules system. 

We explain firstly the codes to simulate the NV centers system. In Fig. \ref{fig:code-NV} (a), line $1$ imports the "QuantumCumulants.jl" and "Plots" packages, and line $2$ and $3$ define the complex variables. Line $4$ defines the Hilbert space of the microwave resonator (as a quantum harmonic oscillator) while line $5$ and $6$ define the Hilbert space for a single NV center (as a 7-level system) and a special Hilbert space for an ensemble of $N$ NV centers, respectively. Note that the last parameter of the $6$th line indicates the number of representative NV center marked by $k = 1,2$. Line $7$ constructs the Hilbert space for the NV centers-microwave resonator system. Line $8$ defines the annihilation operator for photons, and the transition operators ($i\neq j$) and the projection operators ($i=j$) for NV centers and line $9$ defines the system Hamiltonian. Line $10$ and $11$ define the list of operators and rates to specify Lindblad dissipative superoperators in the quantum master equation. Line $12$ defines the list of operators, such as $a{'}a$ for the photon number operator, while line $13$ derives the mean-field equations for the expected value of these operators according to the second-order mean-field approach. Line $14$ analyzes the unknown mean-field quantities and derive the equations for them to finally form a complete set of equations.

In Fig. \ref{fig:code-NV} (b), line $1$ specifies the parameters for the NV centers, and line $2$ computes the optical pumping rate for given laser power. Line $3$ defines the parameters of the microwave resonator. Line $4$ and $5$ define the list of symbols and their values. Line $6$ defines the initial value of mean-field quantities, and line $7$ imports the "OrdinaryDiffEq", "ModelingToolkit" and "SteadyStateDiffEq" packages to define and solve ordinary differential equations (ODE). Line $8$ defines the ODE system and line $9$ defines the ODE problem with the initial value, evolution time and parameters. Line $10$ 
solves the ODE problem with the Runge-Kutta method and line $11$ extracts the evolution time, the intra-resonator photons number and the population of the ground states spin levels.

In Fig. \ref{fig:code-NV} (c), line $1$ calculates the spin coherence, the spin levels populations and the spin-spin correlations, while line $2$ and  $3$ use these results to compute the collective spin vector components, and their uncertainties. Line $4$ calculates the average of the Dicke state quantum numbers, normalized by the number of the NV centers (see below). Here, we utilize the operators $\sigma(1,1)[k]$ ($k=1,2$) to represent the projection operators on the first levels for the two representative NV centers. In practice, the QuantumCumulants.jl package does not include directly these operators, but represents them internally by $1-\sum_{j=2,...,7}\sigma(j,j)[k]$. To calculate the spectrum, line $5$ constructs the correlation function $\left\langle \hat{a}^\dagger(\tau) \hat{a}(0) \right \rangle$, and line $6$ defines the Fourier transform of the correlation function. Then, line $7$ and $8$ define and then solves the steady-state problem of the system. Line $9$ defines the list of frequencies, and computes the spectrum for these frequencies. Line $10$ and $11$ plot the spectrum, and the dynamics of the mean photon number. 

The Julia codes to simulate the pentacene molecules system are shown in Fig. \ref{fig:code-NV} (d-f), and are similar to those for the NV centers system shown in Fig. \ref{fig:code-NV} (a-c). Thus, we do not explain the details here.

In the following, we explain how to compute the average of the Dicke states quantum numbers. We compute first the collective spin vector components, i.e. the expectation values of the operators $\hat{j}^x=(1/2)\sum_k (\left\langle \hat{\sigma}_{k}^{12}\right\rangle+\left\langle \hat{\sigma}_{k}^{21}\right\rangle),\hat{j}^y=(i/2)\sum_k (\left\langle \hat{\sigma}_{k}^{12}\right\rangle-\left\langle \hat{\sigma}_{k}^{21}\right\rangle),\hat{j}^z=(1/2)\sum_k (\left\langle \hat{\sigma}_{k}^{22}\right\rangle-\left\langle \hat{\sigma}_{k}^{11}\right\rangle)$:  $A_{x}=\left\langle \hat{j}^{x}\right\rangle = N/2\left(\left\langle \hat{\sigma}_{1}^{12}\right\rangle +\left\langle \hat{\sigma}_{1}^{21}\right\rangle \right)$, $A_{y}=\left\langle \hat{j}^{y}\right\rangle =iN/2\left(\left\langle \hat{\sigma}_{1}^{12}\right\rangle -\left\langle \hat{\sigma}_{1}^{21}\right\rangle \right)$, and $A_{z}=\left\langle \hat{j}^{z}\right\rangle =N/2\left(\left\langle \hat{\sigma}_{1}^{22}\right\rangle -\left\langle \hat{\sigma}_{1}^{11}\right\rangle\right)$. We then compute the expectation values of the square of the same operators: 

\begin{align}
& B_x = \left\langle \left(\hat{j}^{x}\right)^{2}\right\rangle   =N/4\bigl\{\left(N-1\right)(\left\langle \hat{\sigma}_{1}^{12}\hat{\sigma}_{2}^{12}\right\rangle  + \left\langle \hat{\sigma}_{1}^{12}\hat{\sigma}_{2}^{21}\right\rangle \nonumber \\
&+\left\langle \hat{\sigma}_{1}^{21}\hat{\sigma}_{2}^{12}\right\rangle +  \left\langle \hat{\sigma}_{1}^{21}\hat{\sigma}_{2}^{21}\right\rangle)  + 1 \bigr\},
\end{align}
\begin{align}
& B_y = \left\langle \left(\hat{j}^{y}\right)^{2}\right\rangle   =-N/4\bigl\{\left(N-1\right) [\left\langle \hat{\sigma}_{1}^{12}\hat{\sigma}_{2}^{12}\right\rangle -  \left\langle \hat{\sigma}_{1}^{12}\hat{\sigma}_{2}^{21}\right\rangle  \nonumber \\
&- \left\langle \hat{\sigma}_{1}^{21}\hat{\sigma}_{2}^{12}\right\rangle + \left\langle \hat{\sigma}_{1}^{21}\hat{\sigma}_{2}^{21}\right\rangle ]-1  \bigr\},
\end{align}

\begin{align}
&B_z =  \left\langle \left(\hat{j}^{z}\right)^{2}\right\rangle   = N/4 \{ \left(N-1\right)(\left\langle \hat{\sigma}_{1}^{22}\hat{\sigma}_{2}^{22}\right\rangle -2\left\langle \hat{\sigma}_{1}^{11}\hat{\sigma}_{2}^{22}\right\rangle \nonumber \\
&+ \left\langle \hat{\sigma}_{1}^{11}\hat{\sigma}_{2}^{11}\right\rangle) + (\left\langle \hat{\sigma}_{1}^{11}\right\rangle + \left\langle \hat{\sigma}_{1}^{22}\right\rangle) \}.
\end{align}
From these results, we obtain the average of the Dicke state quantum numbers as  $M = J_z$, $J = \sqrt{B_x + B_y + B_z}$. For the pentacene molecules system, we calculate the average of the Dicke states quantum numbers, related to the $3$ and $5$ spin levels coupled to the microwave resonator.

\section{Extra Numerical Results} \label{sec:extra}

In this Appendix, we provide extra numerical results to the NV centers-microwave resonator system and the pentacene molecules-microwave resonator system.

\begin{figure}[!htp]
\begin{centering}
\includegraphics[scale=0.29]{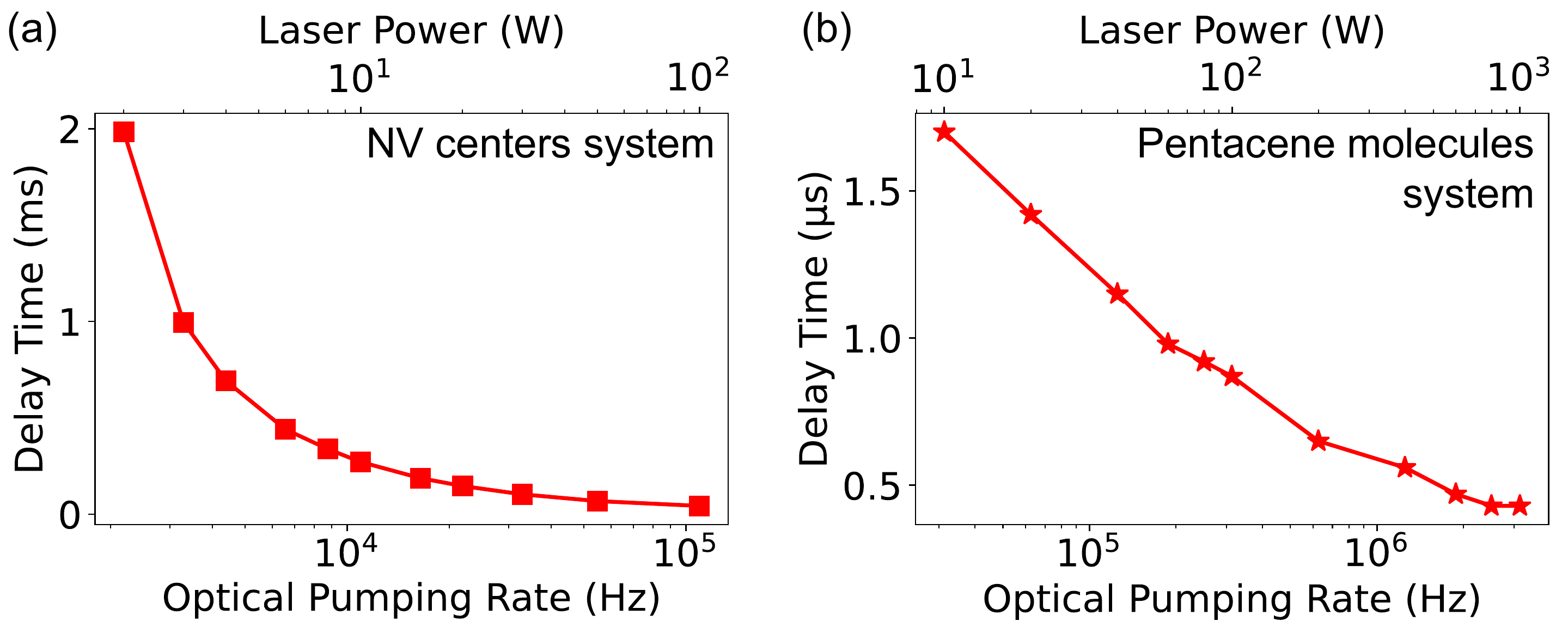}
\par\end{centering}
\caption{\label{fig:start} Delay time of superradiant Rabi oscillations at room temperature as function of optical pumping rate (lower axes) or laser power (upper axes) for the systems with the NV centers (a), and the pentacene molecules (b).} 
\end{figure}

\subsection{Delay Time of Rabi Oscillations \label{se:delay}}

In the main text, we have examined in detail the dependence of the frequency and duration of superradiant Rabi oscillations under the optical pumping at room temperature. In Fig. \ref{fig:start}, we complement these results by analyzing the influence of the optical pumping on the delay time of the onset of the superradiant Rabi oscillations. For the system with NV centers, the delay time decreases gradually from about $2$ ms to almost zero with increasing optical pumping rate (laser power), as shown in Fig. \ref{fig:start} (a). For the system with pentacene molecules, the delay time reduces from $1.7$ $\mu$s to $0.4$ $\mu$s and is about three orders of magnitude faster than in the NV centers system, as shown in Fig. \ref{fig:start} (b).

\begin{figure}[htp]
\begin{centering}
\includegraphics[scale=0.26]{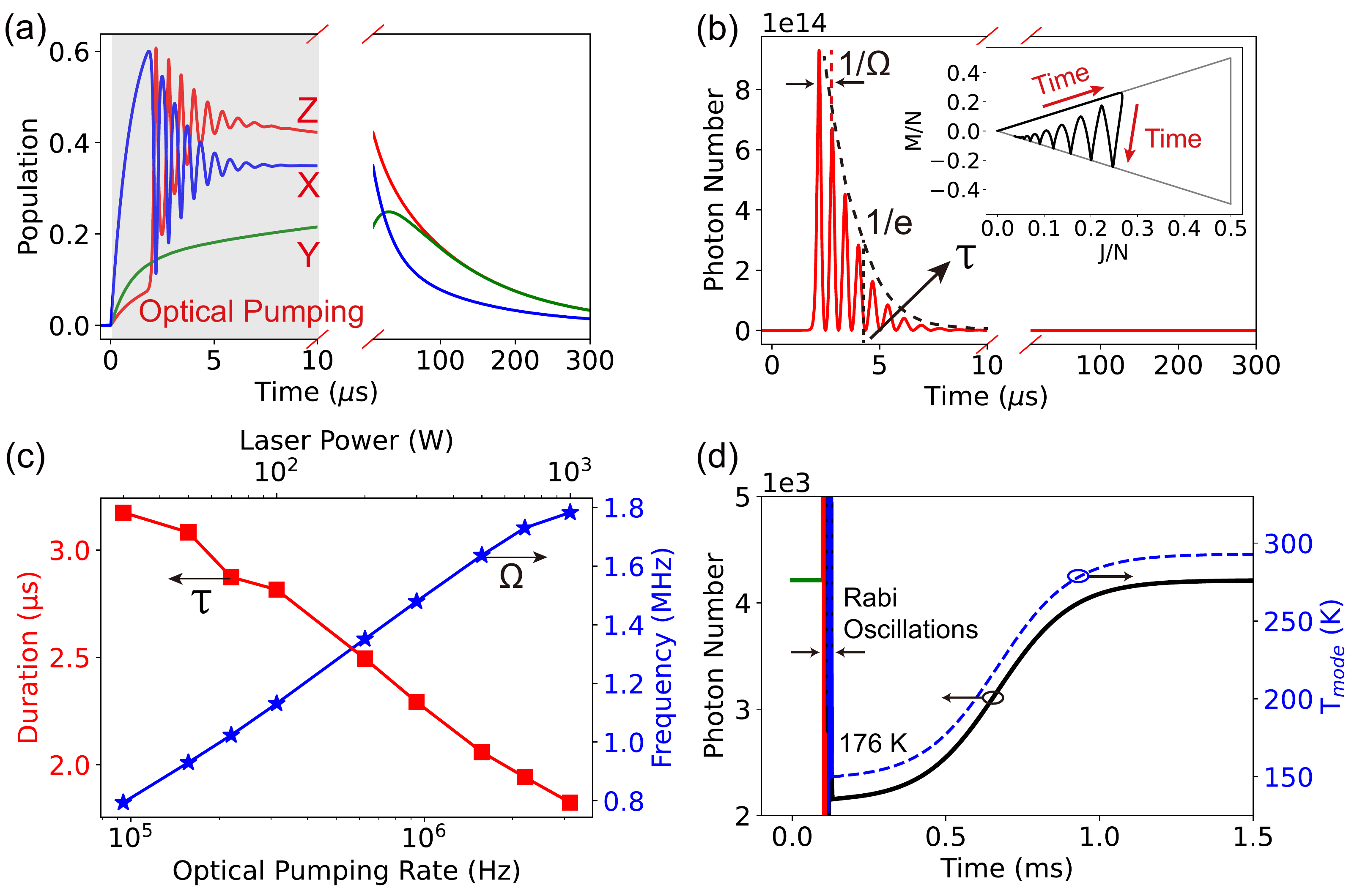}
\par\end{centering}
\caption{\label{fig:rabi-short} Panel (a-c) shows similar results for the pentacene molecules system as Fig. 2 (d-f) in the main text, which reproduce quite well the results in the experiment~\citep{JDBreeze}. Panel (d) shows the dynamics of the intra-resonator photon number and the effective mode temperature for longer time.}
\end{figure}

\subsection{Simulations of Pentacene Molecules System with Different Spin Parameters \label{sub:RabiPulses}}

In the following, we simulate the pentacene molecules system by using the values of the spin decay and dephasing reported in the experiment~\citep{JDBreeze}, see the sixth column of Tab.~\ref{fig:tab}. Here, we observe similar results as shown in Fig. 2 (d-f) in the main text, except that the population of the $X$ spin level is smaller than that of the $Z$ spin level at long time, and as a result, the intra-resonator photon number can not be maintained at high value at long time, i.e. the disappearance of the continuous-wave masing.  In the main text, we emphasize that microwave mode cooling is achieved in the pentacene molecules system when the optical pumping is switched off. This point is clearly illustrated in Fig. \ref{fig:rabi-short} (d). We  find that indeed the photon number reduces down to $2528$ below the thermal value $4211$ after the superradiant Rabi oscillations. In addition, we find that the photon number increases gradually back to the thermal value in about $1.5$ ms. Using the formula $T_{mode}=\hbar\omega_{m}/\left[k_{B}\mathrm{ln}\left(1/\langle\hat{a}^{\dagger}\hat{a}\rangle+1\right)\right]$, we estimate that the effective mode temperature reaches about $176$ K after the optical pumping.


\begin{thebibliography}{1}


\bibitem{RHDicke1954} R. H. Dicke, \textit{Coherence in Spontaneous Radiation
Processes}. Phys. Rev. \textbf{93}, 99 (1954).

\bibitem{AVAndreev1980} A. V. Andreev, V. I. Emel\textquoteright yanov and Y. A. II\textquoteright inskii, \textit{Collective Spontaneous Emission (Dicke Superradiance)}. Sov. Phys. Usp. \textbf{23}, 493 (1980).

\bibitem{DMeiser2009} D. Meiser, J. Ye, D. R. Carlson and M. J. Holland, \textit{Prospects for a Millihertz-Linewidth Laser}. Phys Rev Lett. \textbf{102}, 163601 (2009). 

\bibitem{ADLudlow} A. D. Ludlow, M. M. Boyd, J. Ye, E. Peik and P. O. Schmidt, \textit{Optical atomic clocks}. Rev. Mod. Phys. \textbf{87}, 637-701 (2015).

\bibitem{DATieri2017} D. A. Tieri, M. Xu, D. Meiser, J. Cooper and M. J. Holland, \textit{Theory of the Crossover from Lasing to Steady State Superradiance}, arXiv:1702.04830.

\bibitem{KDebnath2018} K. Debnath, Y. Zhang and K. M{\o}lmer, \textit{Lasing in the Superradiant Crossover Regime}. Phys. Rev. A \textbf{98}, 063837 (2018). 

\bibitem{YZhang2021} Y. Zhang, C. Shan and K. M{\o}lmer, \textit{Ultranarrow Superradiant Lasing by Dark Atom-Photon Dressed States}. Phys. Rev. Lett.  \textbf{126}, 123602 (2021). 

\bibitem{JDBohnet2012} J. G. Bohnet, Z. Chen, J. M. Weiner, D. Meiser, M. J. Holland and J. K. Thompson, \textit{A Steady-state Superradiant Laser with Less
than One Intracavity Photon}. Nature \textbf{484}, 78-81 (2012). 

\bibitem{MANorcia2016} M. A. Norcia, M. N. Winchester, J. R. K. Cline and J. K. Thompson, \textit{Superradiance on the Millihertz Linewidth Strontium Clock Transition}. Sci. Adv. \textbf{2}, e1601231 (2016). 

\bibitem{MANorcia2016-1} M. A. Norcia and J. K. Thompson, \textit{Cold-strontium Laser in the Superradiant Crossover Regime}. Phys. Rev. X \textbf{6}, 011025 (2016). 

\bibitem{MANorcia2018} M. A. Norcia, J. R. K. Cline, J. A. Muniz, J. M. Robinson, R. B. Hutson, A. Goban, G. E. Marti, J. Ye and J. K. Thompson,
\textit{Frequency Measurements of Superradiance from the Strontium Clock Transition}. Phys. Rev. X \textbf{8}, 21036 (2018). 

\bibitem{YZhang3} Y. Zhang, C. Shan and K. M{\o}lmer, \textit{Active Frequency Measurement on Superradiant Strontium Clock Transitions}, Phys. Rev. Lett. \textbf{128}, 013604 (2022). 

\bibitem{TSLin} T.-S. Lin,  \textit{Novel Pulsed Electron Paramagnetic Resonance Techniques for the Studies of Structure and Dynamics of Photo-excited Triplet State of Organic Molecules: A Professional Journey.} J. Chin. Chem. Soc. \textbf{65}, 163-188 (2018). 

\bibitem{BarryJF} J. F. Barry, J. M. Schloss, E. Bauch, M. J. Turner, C. A. Hart, L. M. Pham and R. L. Walsworth. \textit{Sensitivity optimization for NV$^-$ diamond magnetometry.} Rev. Mod. Phys. \textbf{92}, 015004 (2020).

\bibitem{Kohler} J. K{\"o}hler,  J. A. J. M. Disselhorst, M. C. J. M.  Donckers, E. J. J. Groenen, J. Schmidt and W. E. Moerner,  \textit{Magnetic Resonance of a Single Molecular Spin.} Nature, \textbf{363}, 242-244 (1993).


\bibitem{JWrachtrup1993} J. Wrachtrup, C. von Borczyskowski, J. Bernard, M. Orrit and R. Brown,  \textit{Optical Detection of Magnetic Resonance in a Single Molecule.} Nature, \textbf{363}, 244-245 (1993).


\bibitem{AGruber} A. Gruber, A. Dr{\"a} benstedt, C. Tietz, L. Fleury, J. Wrachtrup and C. von Borczyskowski, \textit{Scanning Confocal Optical Microscopy and Magnetic Resonance on Single Defect Centers.} Science, \textbf{276}, 2012-2014 (1997).

\bibitem{Breeze} J. D. Breeze, E. Salvadori, J. Sathian, N. M. N. Alford and C. W. M. Kay, \textit{Continuous-wave Room-temperature Diamond Maser}, Nature \textbf{555}, 493 (2018).

\bibitem{Oxborrow} M. Oxborrow, J. D. Breeze and N. M. Alford, \textit{Room-temperature Solid-state Maser}, Nature \textbf{488}, 353 (2012).

\bibitem{HWu1} H. Wu, X. Xie, W. Ng, S. Mehanna, Y. Li, M. Attwood and M. Oxborrow, \textit{Room-Temperature Quasi-Continuous-Wave Pentacene Maser Pumped by an Invasive Ce:YAG Luminescent Concentrator}, Phys. Rev. Appl. \textbf{14}, 064017 (2020).

\bibitem{ESalvadori} E. Salvadori, J. D. Breeze, K.-J. Tan, J. Sathian, B. Richards, M. W. Fung, G. Wolfowicz, M. Oxborrow, N. M. Alford and C. W. M. Kay, \textit{Nanosecond Time-resolved Characterization of a Pentacene-based Room-temperature Maser}, Sci. Rep. \textbf{7}, 41836 (2017).

\bibitem{AAngerer2018} A. Angerer, K. Streltsov, T. Astner, S. Putz, H. Sumiya, S. Onoda, J. Isoya, W. J. Munro, K. Nemoto, J. Schmiedmayer and J. Majer, \textit{Superradiant Emission from Colour Centres in Diamond.} Nat. Phys. \textbf{14}, 1168-1172 (2018)

\bibitem{SPutz} S. Putz, D. O. Krimer, R. Ams{\"u}ss, A. Valookaran, T. N{\"o}bauer, J. Schmiedmayer, S. Rotter and J. Majer, \textit{Protecting a Spin Ensemble against Decoherence in the Strong-coupling Regime of Cavity QED}, Nat. Phys. \textbf{10}, 720-724 (2014).

\bibitem{JDBreeze} J. D. Breeze, E. Salvadori, J. Sathian, N. M. Alford and C. W. M. Kay, \textit{Room-temperature Cavity Quantum Electrodynamics with Strongly Coupled Dicke States}, npj Quantum Inf. \textbf{3}, 40 (2017).

\bibitem{AAngerer} A. Angerer, T. Astner, D. Wirtitsch, H. Sumiya, S. Onoda, J. Isoya, S. Putz and J. Majer, \textit{Collective Strong Coupling with Homogeneous Rabi Frequencies using a 3D Lumped Element Microwave Resonator}, Appl. Phys. Lett. \textbf{109}, 033508 (2016).

\bibitem{RAmsuss} R. Ams{\"u}ss, C. Koller, T. N{\"o}bauer, S. Putz, S. Rotter, K. Sandner, S. Schneider, M. Schramb{\"o}ck, G. Steinhauser, H. Ritsch, J. Schmiedmayer and J. Majer, \textit{Cavity QED with Magnetically Coupled Collective Spin States}, Phys. Rev. Lett. \textbf{107}, 060502 (2011).

\bibitem{YZhang2022PRL} Y. Zhang, Q. Wu, S.-L. Su, Q. Lou, C. Shan and K. M{\o}lmer, \textit{Cavity Quantum Electrodynamics Effects with Nitrogen Vacancy Center Spins Coupled to Room Temperature Microwave Resonators}, Phys. Rev. Lett. \textbf{128}, 253601 (2022).

\bibitem{DPFahey} D. P. Fahey, K. Jacobs, M. J. Turner, H. Choi, J. E. Hoffman, D. Englund and M. E. Trusheim, \textit{Steady-state Microwave Mode Cooling with a Diamond NV Ensemble}, arXiv:2203:03462v1. 





\bibitem{DPl} D. Plankensteiner, C. Hotter and H. Ritsch, \textit{QuantumCumulants.jl: A Julia Framework for Generalized Mean-field Equations in Open Quantum Systems}, Quantum \textbf{6}, 617 (2022).

\bibitem{CHotter} C. Hotter, D. Plankensteiner, G. Kazakov and H. Ritsch, \textit{Continuous multi-step Pumping of the Optical Clock Transition in Alkaline-Earth Atoms with Minimal Perturbation.} Opt. Express \textbf{30}, 5553-5568 (2022). 

\bibitem{PMeystre} P. Meystre and M. Sargent, Elements of Quantum Optics. Springer, New York, 2007.

\bibitem{HKraus} H. Kraus, V. A. Soltamov,  D. Riedel,  S. V{\"a}th,  F. Fuchs,  A. Sperlich,  P. G. Baranov,  V. Dyakonov and  G. V. Astakhov, \textit{Room-temperature Quantum Microwave Emitters based on Spin Defects in Silicon Carbide.} Nat. Phys. \textbf{10}, 157-162 (2014).

\bibitem{MFischer} M. Fischer, A. Sperlich,  H. Kraus, T. Ohshima, G. V. Astakhov and V. Dyakonov,  \textit{Highly Efficient Optical Pumping of Spin Defects in Silicon Carbide for Stimulated Microwave Emission.} Phys. Rev. Appl. \textbf{9}, 54006 (2018).

\bibitem{AGottscholl2020} A. Gottscholl, M. Kianinia,  V. Soltamov, S. Orlinskii, G. Mamin, C. Bradac, C. Kasper, K. Krambrock, A. Sperlich, M. Toth, I. Aharonovich and V. Dyakonov, \textit{Initialization and Read-out of Intrinsic Spin Defects in a van der Waals Crystal at Room Temperature.} Nat. Mater. \textbf{19}, 540-545  (2020).

\bibitem{AGottscholl2021}  A. Gottscholl,  M. Diez, V. Soltamov,  C. Kasper,  D. Krau{\ss}e,  A. Sperlich,  M. Kianinia,  C. Bradac, I. Aharonovich and V. Dyakonov, \textit{Spin Defects in hBN as Promising Temperature, Pressure and Magnetic Field Quantum Sensors.} Nat. Commun. \textbf{12}, 4480 (2021).

\bibitem{WNg} W. Ng, H. Wu and M. Oxborrow, \textit{Continuous Cooling of a Microwave Mode on a Benchtop using Hyperpolarized NV$^{-}$ Diamond}, Appl. Phys. Lett. \textbf{119}, 234001 (2021).

\bibitem{YZhang1} Y. Zhang, Q. Wu, H. Wu, X. Yang, S.-L. Su, C. Shan and K. M{\o}lmer, \textit{Cavity Quantum Electrodynamics Effects of Optically Cooled Nitrogen-Vacancy Centers Coupled to a High Frequency Microwave Resonator}, npj Quantum Inf. \textbf{8}, 125 (2022).

\bibitem{HWu2019} H. Wu, W. Ng, S. Mirkhanov, A. Amirzhan, S. Nitnara and M. Oxborrow, \textit{Unraveling the Room-Temperature Spin Dynamics of Photoexcited Pentacene in Its Lowest Triplet State at Zero Field}, J. Phys. Chem. C, \textbf{123}, 24275 (2019).

\bibitem{QWu2021} Q. Wu, Y. Zhang, X. Yang, S.-L. Su, C. Shan and K. M{\o}lmer, \textit{A Superradiant Maser with Nitrogen-vacancy Center Spins},  Sci. China: Phys. Mech. Astron. \textbf{65}, 217311 (2022).

\bibitem{YZhang4} Y. Zhang, Y-X. Zhang and K. M{\o}lmer, \textit{Monte-Carlo Simulations of Superradiant Lasing}, New J. Phys. \textbf{20}(11), 112001 (2018).


\bibitem{HWu} H. Wu, S. Mirkhanov, W. Ng and M. Oxborrow, \textit{Bench-Top Cooling of a Microwave Mode Using an Optically Pumped Spin Refrigerator}, Phys. Rev. Lett. \textbf{127}, 053604 (2021).

\bibitem{ASarkar} A. Sarkar, B. Blankenship, E. Druga, A. Pillai, R. Nirodi, S. Singh, A. Oddo, P. Reshetikhin and  A. Ajoy, \textit{Rapidly Enhanced Spin-Polarization Injection in an Optically Pumped Spin Ratchet.} Phys. Rev. Appl. \textbf{18}, 34079 (2022).

\bibitem{CSzczuka} C. Szczuka, M. Drake and J. A.  Reimer, \textit{Effects of laser-induced heating on nitrogen-vacancy centers and single-nitrogen defects in diamond.} J. Phys. D \textbf{50}, 395307 (2017). 

\bibitem{AJarmola}  A. Jarmola, V. M. Acosta, K. Jensen, S. Chemerisov and  D. Budker, \textit{Temperature- and Magnetic-Field-Dependent Longitudinal Spin Relaxation in Nitrogen-Vacancy Ensembles in Diamond.} Phys. Rev. Lett. \textbf{108}, 197601 (2012).

\bibitem{VMAcosta} V. M. Acosta, E. Bauch, M. P. Ledbetter, A. Waxman, L.-S. Bouchard and  D. Budker, \textit{Temperature Dependence of the Nitrogen-Vacancy Magnetic Resonance in Diamond.} Phys. Rev. Lett. \textbf{104}, (2010)

\bibitem{XDChen} X. D. Chen, C. L. Zou, F. W. Sun and G. C. Guo, \textit{Optical Manipulation of the Charge State of Nitrogen-vacancy Center in Diamond.} Appl. Phys. Lett. \textbf{103}, 13112 (2013). 


\bibitem{THolstein} T. Holstein and H. Primakoff, \textit{Field Dependence of the Intrinsic Domain Magnetization of a Ferromagnet}, Phys. Rev. \textbf{58}, 1098 (1940)





\end{thebibliography}
\end{document}